\documentclass[lettersize,journal]{IEEEtran}
\usepackage{graphicx}
\usepackage{cite}
\usepackage{picinpar}
\usepackage{amsmath}
\usepackage{amssymb}
\usepackage{amsfonts}
\usepackage{amsthm}
\usepackage{url}
\usepackage{flushend}
\usepackage[latin1]{inputenc}
\usepackage{colortbl}
\usepackage{soul}
\usepackage{multirow}
\usepackage{pifont}
\usepackage{color}
\usepackage{alltt}
\usepackage[hidelinks]{hyperref}
\usepackage{enumerate}
\usepackage{siunitx}
\usepackage{breakurl}
\usepackage{epstopdf}
\usepackage{pbox}
\usepackage[caption=false,font=normalsize,labelfont=sf,textfont=sf]{subfig}
\usepackage{caption}
\usepackage{diagbox}

\newenvironment{Note to Practitioners}
{
	\renewcommand{\abstractname}{Note to Practitioners}
	\begin{abstract}
	}
	{
	\end{abstract}
}

\newtheoremstyle{nonbold}
{}                        
{}                        
{}                		  
{}                        
{}                        
{\textit:}                
{ }                       
{\textit{\thmname{#1}\thmnumber{ #2}\thmnote{ (#3)}}} 

\theoremstyle{nonbold}
\newtheorem{Theorem}{Theorem}
\newtheorem{Proof}{Proof of Theorem}
\newtheorem{Assumption}{Assumption}

\begin{document}
	
	\title{Compact Amplified Laser Power Stabilization Using Robust Active Disturbance Rejection Control with Sensor Noise Decoupling}
	
	\author{
		\vskip 1em
		
		Yanpei Shi, Jingxuan Zhang, Zhuo Shi, Chenyao Zhang, Yuze Guo, and Rui Feng
		
		\thanks{
			
			
		 		Yanpei Shi, Zhuo Shi, Chenyao Zhang, Yuze Guo, and Rui Feng are affiliated with School of Instrumentation and Optoelectronic Engineering, Beihang University, Beijing 100191, China (e-mail: \url{yanpei_shi@buaa.edu.cn}; \url{fengrui_buaa@buaa.edu.cn}).
			
		 		Jingxuan Zhang is affiliated with the Yau Mathematical Sciences Center, Tsinghua University, Beijing 100084, China.
		 	}
	}
	
	\maketitle
	
	\begin{abstract}
		Laser power instability, encompassing random jitter and slow drift, severely limits the performance of optically pumped magnetometers (OPMs) in detecting ultra-weak magnetic fields, especially in large-scale OPM arrays for magnetoencephalography. Although a unified amplified laser (AL) architecture improves integration, fluctuations in the pump beam progressively degrade performance across all channels, exacerbated by environmental disturbances and system uncertainties. To address this challenge, this paper presents a compact AL power stabilization approach based on an innovative dual-loop active disturbance rejection control (DLADRC) strategy, while integrating a comprehensive quantitative stability analysis through novel exponential decay estimates for extended state observers (ESOs) and control error dynamics. As validated through physical experimental results, the proposed method significantly improves AL's long-term stability with sensor noise decoupling, achieving an over 85.7\% reduction in 1-hour power instability and a tenfold decrease in Allan variance for correlation times 10\textsuperscript{2} s--10\textsuperscript{3} s, compared to standard ADRC. Crucially, the strategy demonstrates robust effectiveness across diverse operating scenarios, enabling AL-based OPM systems to achieve their full potential in high-sensitivity biomagnetic field detection.
	\end{abstract}
	
	\begin{Note to Practitioners}
		This article focuses on AL power control with compact structure and proposes a direct feedback architecture with a novel DLADRC control strategy, which is
		capable of achieving AL power stabilization  subject to sensor noise and unknown disturbances with only a single photodetector. The article contains a comprehensive methodology encompassing system modeling, control strategy design, quantitative stability analysis, and experimental validation, thereby offering practitioners a clear reference framework for implementing similar control systems. 
		On the theoretical side, a unified estimation framework based on Lyapunov functionals is provided to derive time-domain explicit stability criteria and further analyze the transient and steady-state performance of DLADRC systems through quantitative estimates on observation and control errors.	The proposed control strategy and analysis method can be readily extended to a broad class of ADRC-based systems, thus facilitating practitioners with a practical framework for robust control of industrial noise-coupled systems operating under diverse conditions. Finally, benchmark tests validate the theoretical predictions and reveal the proposed method's superior performance compared to conventional ADRC structure, empirically underscoring its viability for practical deployment in analogous systems.
	\end{Note to Practitioners}
	
	
	
	\definecolor{limegreen}{rgb}{0.2, 0.8, 0.2}
	\definecolor{forestgreen}{rgb}{0.13, 0.55, 0.13}
	\definecolor{greenhtml}{rgb}{0.0, 0.5, 0.0}
	
	\section{Introduction}
	\label{sec.1}
	
	\IEEEPARstart{L}{arge-scale} optically pumped magnetometer (OPM) arrays revolutionize magnetoencephalography technology, empowering breakthroughs in both cognitive neuroscience research and clinical psychiatric diagnostics\cite{brookes2022,borna2019,boto2018}. In response to the dual requirements of power gain and reliable integration, a unified amplified laser (AL) architecture is implemented to deliver watt-level pump beams for hundreds-channel OPM arrays, thereby facilitating high-resolution magnetic source reconstruction\cite{rea2022,liu2023}. However, during long-term operation, power fluctuations in the pump laser critically degrade the OPM arrays performance across all channels, stemming from the fundamental coupling in OPMs between light intensity variations and measured biomagnetic signals\cite{colombo2016,long2024}. Consequently, ultra-stable lasers with broad power ranges become an essential prerequisite for sustaining diagnostic-grade reliability of AL-based OPM systems.
	
	The scope of this work belongs to laser power stabilization, with emphasis on addressing the challenges in compact AL systems. Researchers have developed various techniques through active control mechanisms and passive physical phenomena to address diverse noise profiles. Active stabilization techniques predominantly employ dynamic feedback systems utilizing electro-optic or acousto-optic modulators to correct power fluctuations in real time\cite{durak2007,nery2021,zhou2023,niu2024}. These systems often integrate multi-parameter regulation, such as digital servo controllers coordinating temperature stabilization and intensity modulation, or radiation pressure sensing coupled with interferometer displacement monitoring the supression of noise across wide frequency ranges\cite{nery2021,zhou2023}. Passive stabilization leverages inherent optical interactions, including optomechanical coupling via optical spring effects and nonlinear processes in cavity-enhanced harmonic generation, which intrinsically dampen power noise without external control loops through energy transfer mechanisms\cite{lee2016,cullen2022,jiao2024}. However, existing methodologies exhibit two primary limitations: 1) both active and passive methods require supplementary optical components---active systems depend on modulators, while passive implementations need specialized cavities or nonlinear crystals, increasing system complexity and alignment challenges; 2) conventional techniques predominantly operate effectively within constrained power regimes, with active approaches restricted by modulator saturation effects and passive techniques limited by material damage thresholds. These challenges underscore the necessity for compact architectures capable of maintaining stability across broader power regimes without auxiliary optics. 
	
	In the view of the above issues, this paper studies a single-photodetector direct feedback architecture with a novel active disturbance rejection control (ADRC)-based strategy, offering a promising approach for compact AL power stabilization. Within the ADRC framework, the concept of generalized disturbance is introduced to comprehensively characterize both external perturbations from the varying environment and inherent uncertainties in system modeling\cite{han2009,sariyildiz2020,sun2023,aliamooei-lakeh2024}. By continuously monitoring the system output through sensors, the extended state observer (ESO) dynamically incorporates real-time estimates of generalized disturbance into its state vector as a supplementary dimension\cite{khalil2014,alonge2017,li2024,chen2025a,gao2025b}. This extended estimate capability facilitates the straightforward implementation of controllers with disturbance rejection mechanisms\cite{feng2020,wu2019}. Nevertheless, the standard single-ESO ADRC architecture exhibits an unavoidable performance trade-off, that is, enhanced tracking accuracy necessitates higher observer bandwidth, conversely exacerbating sensor noise sensitivity\cite{prasov2013,sariyildiz2020,chen2021}. To address these challenges in the compact AL power stabilization system, here we propose a dual-loop ADRC (DLADRC) architecture in the control strategy, and we show both by quantitative analysis and by experimental validation that the proposed strategy maintains sufficient robustness across varying operations, successfully compensating for both unpredictable disturbances and uncertain dynamics.
	
	\begin{figure}[!t]
		\centering
		\includegraphics[width=0.9\linewidth]{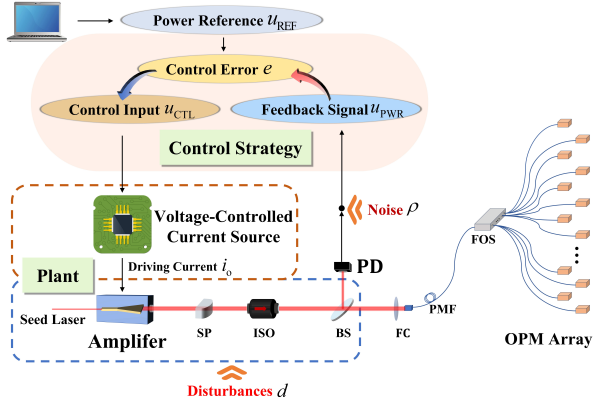}
		\caption{Topology of the compact AL power stabilization system with a single photodetector. PD: photodetector; SP: shaping prism; ISO: isolator; BS: beam splitter; FC: fiber coupler; PMF: polarization-maintaining fiber; FOS: fiber optic splitter.}\label{fig.1}
		\vspace{-3mm}
	\end{figure}
	
	Main contributions of this article are summarized as follows.
	\begin{enumerate}[1)]
		\item We propose a novel DLADRC architecture, with physical experimental demonstration in active AL power stabilization. The inner loop provides independent compensations for system model's mismatches through a supplemental ESO, while the outer loop decouples the photoelectric noise through cascade configuration, as detailed in Section \ref{sec.3a}.
		\item We derive a family of explicit time decay estimates with tractable parameter dependency to characterize the transient and steady-state performance of interconnected ESOs, thereby establishing a generic framework for the stability analysis of DLADRC systems through quantitative estimates on observation and control errors, as detailed in Section \ref{sec.3b} and \ref{sec.3c}.
		\item The simplified physical structure of our method, requiring merely a single photodetector, enables highly compact and reliable employment of ultra-stable lasers. This provides a critical advantage over conventional optical stabilization systems, which are bulky and impractical for clinical applications. Beyond matching optical approaches, our DLADRC-based compact stabilization system outperforms traditional ADRC methods, achieving an 85.7\% lower 1-hour power instability and a tenfold better Allan variance across $10^2\:\mathrm{s}$--$10^3\:\mathrm{s}$ timescales throughout the $1 \text{ W}$--$2 \text{ W}$ range, as detailed in Section \ref{sec.4b}.
		
	\end{enumerate}
	
	This article is organized as follows. In Section \ref{sec.2}, the plant model of AL system is narrated. In Section \ref{sec.3}, the DLADRC strategy is introduced, with a detailed quantitative analysis. In Section \ref{sec.4}, experimental validations are discussed. Finally, Section \ref{sec.5} concludes this article.
	
	\section{Problem Formulation}
	\label{sec.2}
	\subsection{Plant Model and Control Objective}
	\label{sec.2a}
	
	Fig. \ref{fig.1} shows the topology of the proposed compact AL power stabilization system employing a single photodetector to construct direct feedback. Between the control input \( {u_{\rm{CTL}}} \) to the feedback signal \( {u_{\rm{PWR}}} \), the closed-loop system's plant comprises two serial stages: a voltage-controlled current source for amplifier excitation and an optical processing chain for laser gain modulation and shape regulation. 
	
	\begin{figure}[!t]
		\centering
		\subfloat[(a)]
		{		
			\label{fig.2a}
			\begin{minipage}[h]{0.6\linewidth}
				\includegraphics[width=\textwidth]{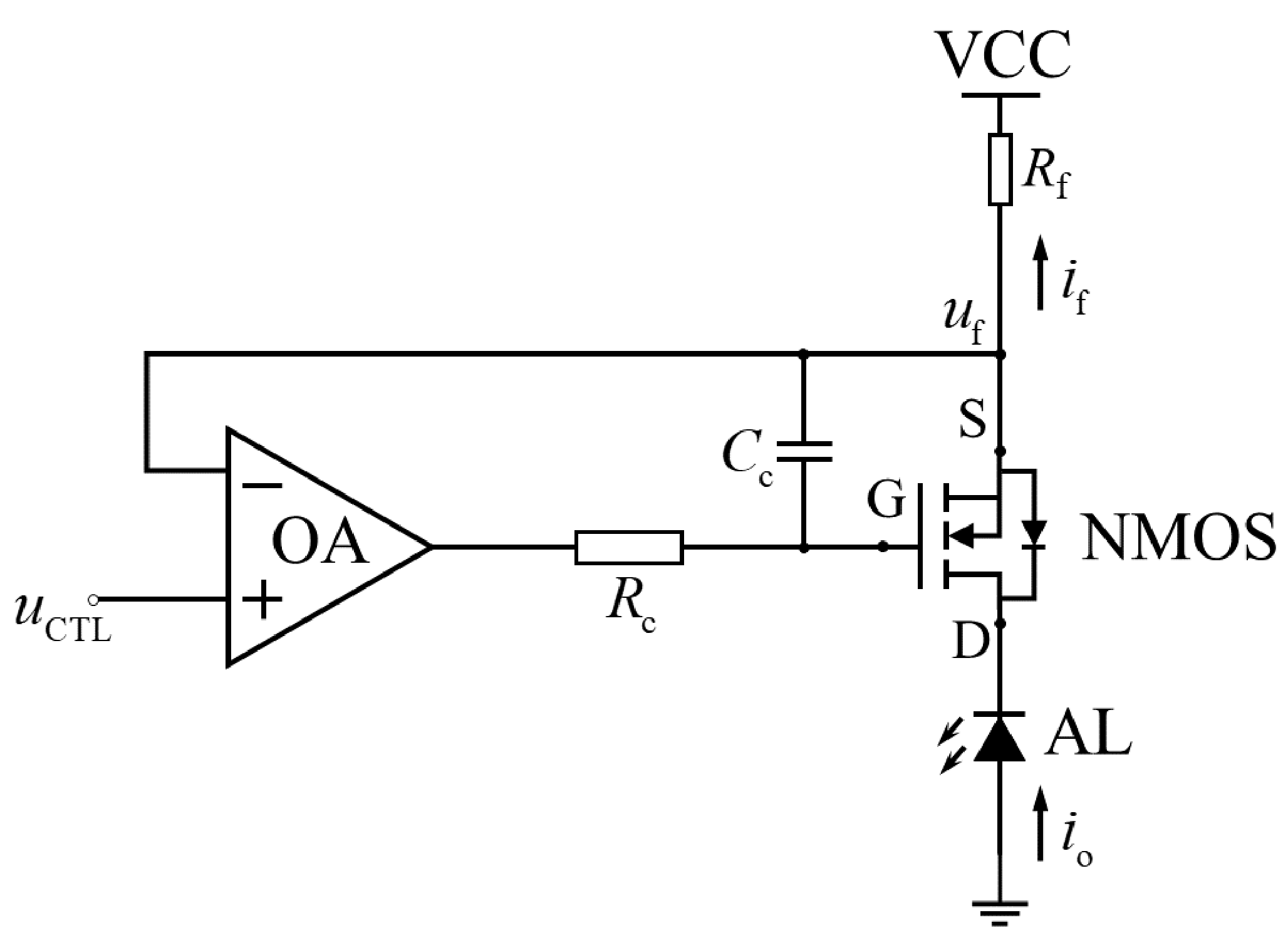}
			\end{minipage}
		}\
		\subfloat[(b)]
		{
			\label{fig.2b}
			\begin{minipage}[h]{0.9\linewidth}
				\includegraphics[width=\textwidth]{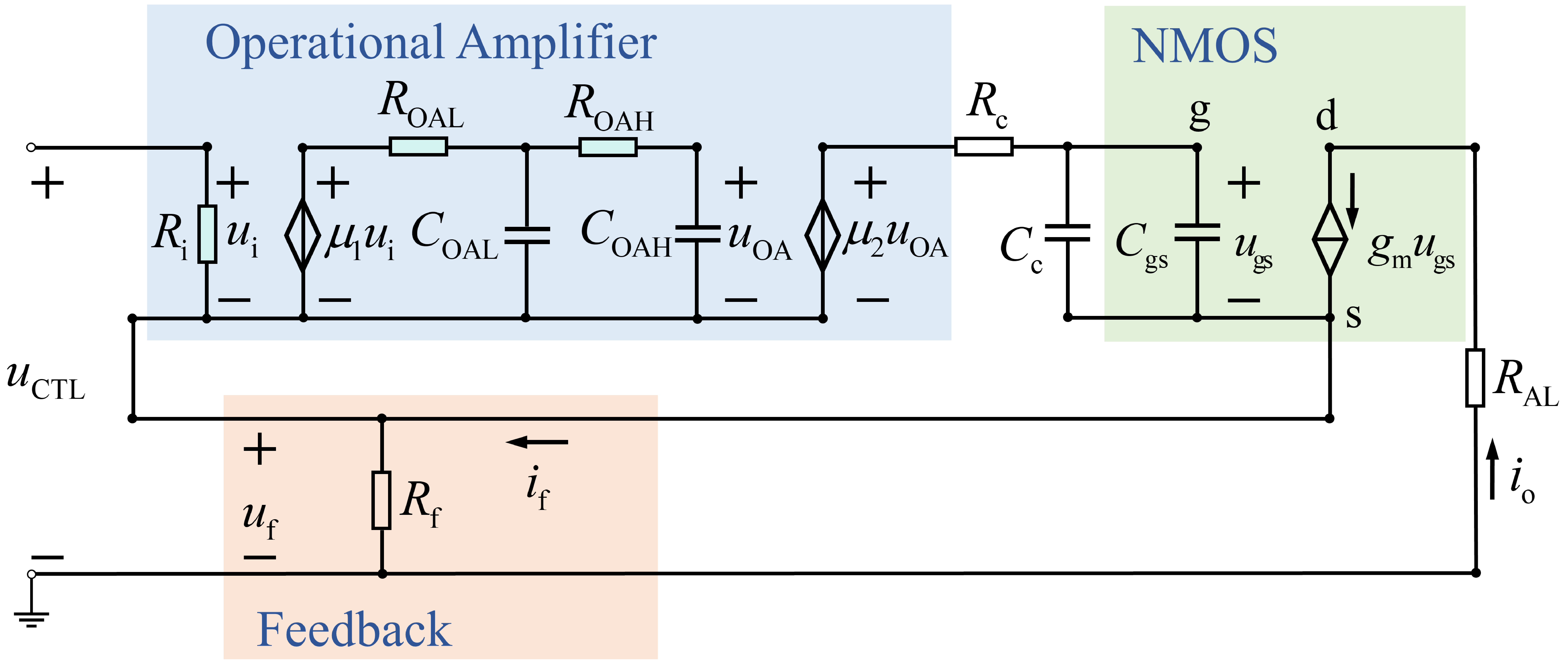}
			\end{minipage}
		}
		
		\caption{Voltage-controlled current source for driving AL. (a) Circuit schematic. (b) High-frequency equivalent circuit model. NMOS: N-channel MOSFET.}
		\label{fig.2}
		\vspace{-6mm}
	\end{figure}
	The system order is determined by the current source constrained by the frequency responses of electronic components, as illustrated in the schematic of Fig. \ref{fig.2}\subref{fig.2a}. By incorporating the hybrid $\pi$ model and accounting for parasitic effects, we develop a comprehensive high-frequency equivalent circuit model in Fig. \ref{fig.2}\subref{fig.2b} to analyze the dynamic of the current source. In contrast, the optical subsystem emerges as the primary susceptibility point wherein external disturbances $d$ induce AL power stability degradation. Owing to the instantaneous nature of optical amplification, the output ${u_{{\rm{PWR}}}}$ relates linearly to the driving current ${i_{\rm{o}}}$ via a constant coefficient ${\phi _{{\rm{AL}}}}$. Therefore, the AL's dynamic model, incorporating both disturbances and measurement noise, is derived as
	\begin{equation} \label{eq.1}
		\begin{aligned}
			\frac{{{d^3}{u_{{\text{PWR}}}}\left( t \right)}}{{d{t^3}}} =  &- \frac{{{\alpha _2}}}{{{\alpha _3}}}\frac{{{d^2}{u_{{\text{PWR}}}}\left( t \right)}}{{d{t^2}}} - \frac{{{\alpha _1}}}{{{\alpha _3}}}\frac{{d{u_{{\text{PWR}}}}\left( t \right)}}{{dt}} \\
			&- \frac{{{\alpha _0}}}{{{\alpha _3}}}{u_{{\text{PWR}}}}\left( t \right) + \frac{\beta }{{{\alpha _3}}}\left[ {{u_{{\text{CTL}}}}\left( t \right) + d\left( t \right)} \right]
		\end{aligned}
	\end{equation}
	where 
	\begin{equation} \label{eq.2}
		\left\{
		\begin{aligned}
			{\alpha _0} &= {\mu _1}{\mu _2}{g_{\rm{m}}}{R_{\rm{f}}} + 1\\
			{\alpha _1} &= {R_{{\rm{OAL}}}}{C_{{\rm{OAL}}}} + {R_{{\rm{OAH}}}}{C_{{\rm{OAH}}}} + {R_{\rm{c}}}\frac{{{C_{{\text{gs}}}}{C_{\text{c}}}}}{{{C_{{\text{gs}}}} + {C_{\text{c}}}}}\\
			{\alpha _2} &= {R_{{\rm{OAL}}}}{C_{{\rm{OAL}}}}{R_{{\rm{OAH}}}}{C_{{\rm{OAH}}}} + {R_{{\rm{OAH}}}}{C_{{\rm{OAH}}}}{R_{\rm{c}}}\frac{{{C_{{\text{gs}}}}{C_{\text{c}}}}}{{{C_{{\text{gs}}}} + {C_{\text{c}}}}} \\
			&+{R_{{\rm{OAL}}}}{C_{{\rm{OAL}}}}{R_{\rm{c}}}\frac{{{C_{{\text{gs}}}}{C_{\text{c}}}}}{{{C_{{\text{gs}}}} + {C_{\text{c}}}}}\\
			{\alpha _3} &= {R_{{\rm{OAL}}}}{C_{{\rm{OAL}}}}{R_{{\rm{OAH}}}}{C_{{\rm{OAH}}}}{R_{\rm{c}}}\frac{{{C_{{\text{gs}}}}{C_{\text{c}}}}}{{{C_{{\text{gs}}}} + {C_{\text{c}}}}}\\
			\beta  &= {\phi _{{\text{AL}}}}{\mu _1}{\mu _2}{g_{\text{m}}}
		\end{aligned}
		\right..
	\end{equation}
	
	The control objective is to regulate the control input \( {u_{{\rm{CTL}}}}\left( t \right) \) so that the AL power indication \( {u_{{\rm{PWR}}}}\left( t \right) \) closely tracks the set point \( {u_{{\rm{REF}}}}\left( t \right) \) in the full operating range of the AL. However, the single-photodetector scheme encounters two primary control challenges: 
	\begin{enumerate}[1)]
		\item Model inaccuracy induced by internal system parameter variations across different power operations;
		\item Stability degradation due to susceptibility to ambient light noise \( \rho \) during high-bandwidth precision measurements.
	\end{enumerate}
	Given AL's implementation in OPM arrays and based on practical operational settings in \cite{nery2021,zhou2023,niu2024,sariyildiz2020,khalil2014,alonge2017,feng2020}, the control system is developed under the following assumption:
	\begin{Assumption} \label{ass.1}
		All values of physical quantities in the AL system, including the aforementioned external disturbances $d\left( t \right)$ and system states, are bounded and continuously differentiable.
	\end{Assumption}	
	
	\subsection{Preliminaries for the DLADRC}
	\label{sec.2b}
	By standard ADRC theory, the generalized disturbance $F$ can be derived from the external disturbances $d$ and the model uncertainties\cite{han2009,gao2003}.
	To clarify the theoretical generality, we extend the third-order system in \eqref{eq.1} to $n$-dimensions and decompose the system's state into inner and outer loops. 
	
	The inner loop investigates the effects of the generalized disturbance $F$ stemming from uncertain model parameters. Accordingly, the inner extended state vector is defined as $\boldsymbol{x} := \left[ \begin{array}{*{20}{c}}{{y_{{\rm{in}}}}}&{{{\dot y}_{{\rm{in}}}}}& \cdots &{y_{{\rm{in}}}^{\left( {n - 1} \right)}}&F\end{array} \right]^ \top $, with the corresponding state-space equations
	\begin{equation} \label{eq.3}
		\left\{
		\begin{aligned}
			\boldsymbol{\dot x} &= \boldsymbol{Ax} + \boldsymbol{d}{{\hat b}_{{\rm{in}}}}{u_{{\rm{in}}}} + \boldsymbol{b}\dot F\\
			y_{{\rm{in}}} &= {\boldsymbol{c}^ \top }\boldsymbol{x}	
		\end{aligned}
		\right.,
	\end{equation}
	where ${u_{{\rm{in}}}}: = {u_{{\rm{CTL}}}}$ is the control input, \({\hat b_{{\rm{in}}}} \) represents a precise-enough estimate of the input gain ${\beta \mathord{\left/{\vphantom {\beta {{\alpha _3}}}} \right.\kern-\nulldelimiterspace} {{\alpha _3}}}$ from \eqref{eq.1}, ${y_{{\rm{in}}}}: = {u_{{\rm{PWR}}}}$ donates the noise-free plant output, $r: = {u_{{\rm{REF}}}}$ is the set point, while the coefficients for ADRC application are given by $\boldsymbol{A} := \left[ {\begin{array}{*{20}{c}}
			{{\boldsymbol{0}^{n \times 1}}}&{{\boldsymbol{I}^{n \times n}}} \\ 
			0&{{\boldsymbol{0}^{1 \times n}}} 
	\end{array}} \right]$, $\boldsymbol{b} := \left[\begin{array}{*{20}{c}} 0 &\cdots &0 &1 \end{array}\right]_{\left(n + 1\right) \times 1}^ \top $, 
	$\boldsymbol{c} := \left[\begin{array}{*{20}{c}} 1 &0 &\cdots &0 \end{array}\right]_{\left(n + 1\right) \times 1}^ \top $, 
	and 
	$\boldsymbol{d}: = \left[\begin{array}{*{20}{c}} 0 & \cdots &0&1&0\end{array}\right]_{\left(n + 1\right) \times 1}^ \top $. 
	
	In contrast, by treating the disturbance-compensated inner loop as a nominal plant, the outer loop specifically addresses the decoupling of sensor noise contamination $\rho $ from the output measurement. Defined in the error domain with $e: = r - {y_{{\rm{in}}}}$, the outer loop's state is constructed as $\boldsymbol{z}: = {\left[ {\begin{array}{*{20}{c}} e&{\dot e}& \cdots &{{e^{(n - 1)}}}&0 \end{array}} \right]^ \top }$, governed by the equations:
	\begin{equation} \label{eq.4}
		\left\{
		\begin{aligned}
			\boldsymbol{\dot z} &= \boldsymbol{Az} - \boldsymbol{d}{u_{{\rm{out}}}}\\
			{y_{{\rm{out}}}} &= e + \rho 
		\end{aligned}
		\right.,
	\end{equation}
	where ${u_{{\rm{out}}}}$ is the output of the outer loop.
	
	\section{Main Result}
	\label{sec.3}
	\subsection{DLADRC design}
	\label{sec.3a}
	\begin{figure}[!t]
		\centering
		\includegraphics[width=0.9\linewidth]{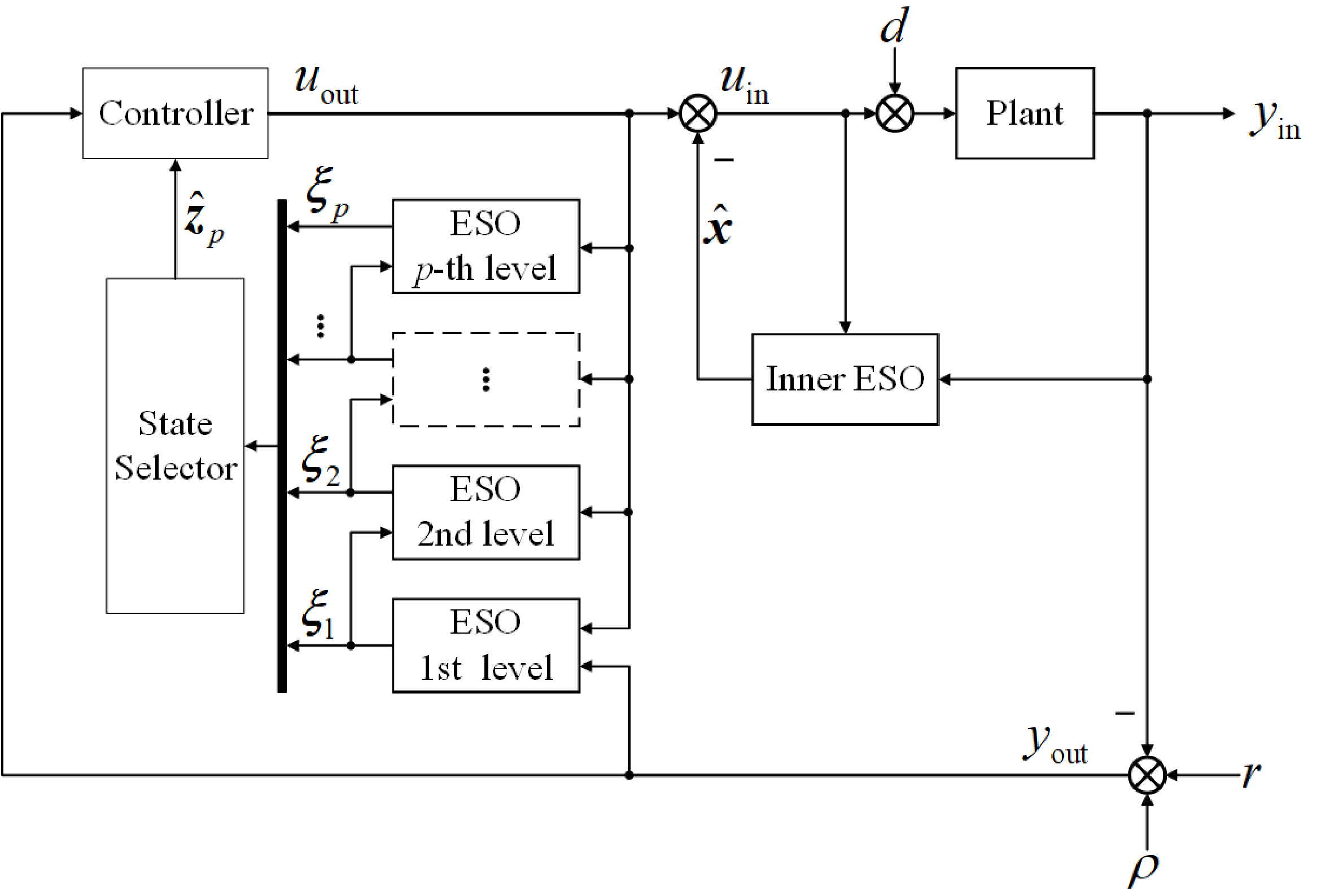}
		\caption{The proposed DLADRC strategy.}\label{fig.3}
	\end{figure}
	
	The proposed DLADRC exploits the twofold benefits from both the nested loops and the cascade topology:
	\begin{enumerate}[1)]
		\item The inner loop's primary role is to counteract uncertain changes of model internal parameters during operational changes, rather than merely suppressing external disturbances. By rapidly adjusting local actuators, it compensates for unmodeled dynamics and maintains tracking accuracy through an isolating mechanism of parameter uncertainties based on the independently tuned ESO, thereby enabling robust control across varying conditions.
		\item In the cascade outer loop structure, the first-level ESO employs a low bandwidth, acting as a low-pass filter to attenuate the measurement noise before it propagates downstream. From the second level onward, each subsequent ESO adopts incrementally higher bandwidths, compensating for the preceding ESO's estimation residuals. This hierarchical design achieves progressive noise suppression and refined state reconstruction.
	\end{enumerate}
	
	Fig. \ref{fig.3} illustrates the architecture of the DLADRC strategy, composed of the following three components.
	\subsubsection{Inner Supplemental ESO}
	To enhance robustness against model mismatches, a independently adjustable linear ESO is embedded into the inner loop. The evolution of inner-loop ESO's observation vector $\boldsymbol{\hat x} \in \mathbb{R}^{n+1}$ is formulated as follows:
	\begin{equation} \label{eq.5}
		\boldsymbol{\dot{\hat{x}}} = \boldsymbol{A\hat x} + \boldsymbol{d}{\hat b_{{\rm{in}}}}{u_{{\rm{in}}}} + \boldsymbol{L}\left( {{y_{{\rm{in}}}} - {\boldsymbol{c}^ \top }\boldsymbol{\hat x}} \right),
	\end{equation}
	where the inner-loop ESO's gain $\boldsymbol{L}$ is specialized as
	\begin{equation} \label{eq.6}
		\begin{aligned}
			\boldsymbol{L} &:= \left[
			\begin{array}{*{20}{c}}
				{{\gamma _1} \times {\omega _{{\rm{in}}}}}
				&{{\gamma _2} \times \omega _{{\rm{in}}}^2}
				& \cdots 
				&{{\gamma _{n + 1}} \times \omega _{{\rm{in}}}^{n + 1}}
			\end{array}\right]^ \top, \\
			{\gamma _m} &:= \frac{{(n + 1)!}}{{m!(n + 1 - m)!}},\quad m \in \left\{ {1, \ldots ,n + 1} \right\},
		\end{aligned}
	\end{equation}
	with a tuning parameter ${\omega _{{\rm{in}}}}$ (${\omega _{{\rm{in}}}} > 1$) donating bandwidth\cite{gao2003,zhou2019}. 
	In practice,  the bandwidths here and below are tuned by the specific experimental setup; see Section \ref{sec.4a} for details. With respect to $\boldsymbol{x}$ in \eqref{eq.3}, the inner observation error is defined as $\boldsymbol{\tilde x}: = \boldsymbol{x} - \boldsymbol{\hat x}$.
	\subsubsection{Outer Cascade ESOs}
	The outer-loop ESOs are structured in a cascade configuration to optimize the noise decoupling:
	\begin{equation} \label{eq.7}
		\left\{
		\begin{aligned}
			{\boldsymbol{\dot \xi }_1} &  = \boldsymbol{A}{\boldsymbol{\xi} _1} - \boldsymbol{d}{u_{{\rm{out}}}} + {\boldsymbol{l}_1}\left( {{y_{{\rm{out}}}} - {\boldsymbol{c}^ \top }{\boldsymbol{\xi} _1}} \right)\\
			{\boldsymbol{\dot \xi }_i} &  = \boldsymbol{A}{\boldsymbol{\xi} _i} + \boldsymbol{d}\left( { - {u_{{\rm{out}}}} + {\boldsymbol{b}^ \top }\mathop \sum \limits_{k = 1}^{i - 1} {\boldsymbol{\xi} _k}} \right) + {\boldsymbol{l}_i}{\boldsymbol{c}^ \top }\left( {{\boldsymbol{\xi} _{i - 1}} - {\boldsymbol{\xi} _i}} \right)\\
			&\hspace{59mm} i \in \left\{ {2, \ldots ,p} \right\}\\
			{\boldsymbol{\hat z}_p} &  = {\boldsymbol{\xi} _p} + \boldsymbol{b}{\boldsymbol{b}^ \top }\sum\limits_{k = 1}^{p - 1} {{\boldsymbol{\xi} _k}} 	 
		\end{aligned}
		\right..
	\end{equation}
	Here, the outer loop's observation ${\boldsymbol{\hat z}_p}\in \mathbb{R}^{n+1}$ consists of the \(p\)-level ESOs, the \(j\)th level ESO's state is 
	\({\boldsymbol{\xi} _j}\in \mathbb{R}^{n+1}\), \({\rm{ }}j \in \left\{ {1, \ldots ,p} \right\}\), and the gain  ${\boldsymbol{l}_j}$ applied to each ESO is given by
	\begin{equation}
		{\boldsymbol{l}_j}: = 
		{[\begin{array}{*{20}{c}}
				{{\gamma _1} \times {\omega _{oj}}}
				&{{\gamma _2} \times \omega _{oj}^2}
				& \cdots 
				&{{\gamma _{n + 1}} \times \omega _{oj}^{n + 1}}
			\end{array}]^ \top }.
	\end{equation}
	By convention, a summation is dropped if its upper limit is smaller than its lower limit. In the cascade outer loop, the bandwidths of the \(p\)-level ESOs form a geometric sequence, i.e., ${\omega _{oj}} = {\alpha ^{j - 1}}{\omega _{o1}}$, for some $\alpha  > 1$ and ${\omega _{o1}} > 1$. The outer observation error is defined as ${\boldsymbol{\tilde z}_p}: = \boldsymbol{z} - {\boldsymbol{\hat z}_p}$, with respect to $\boldsymbol{z}$ in \eqref{eq.4},
	\subsubsection{Controller}
	The control law regulating ${u_{{\rm{in}}}}$ and ${u_{{\rm{out}}}}$ is a linear combination of ${y_{{\rm{out}}}}$, $\boldsymbol{\hat x}$ and $\boldsymbol{\hat z}_p$, with the expressions:
	\begin{equation}
		\left\{
		\begin{aligned}
			{u_{{\rm{out}}}} &= {{\kappa _1} \times \omega _{\rm{c}}^n{y_{{\rm{out}}}} + {\boldsymbol{K}_{{\rm{out}}}}{\boldsymbol{\hat z}_p}}\\
			{u_{{\rm{in}}}} &= \hat b_{{\rm{in}}}^{ - 1}\left( {{u_{{\rm{out}}}} - \boldsymbol{b\hat x}} \right)		
		\end{aligned}
		\right.,
	\end{equation}
	where ${\boldsymbol{K}}_{\rm{out}}$ is parameterized by the control bandwidth ${\omega}_{\rm{c}}$ as
	\begin{equation}
		\begin{aligned}
			{\boldsymbol{K}_{{\rm{out}}}}&: = \left[\begin{array}{*{20}{c}}
				0&{{\kappa _2} \times \omega _c^{n - 1}}& \cdots &{{\kappa _n} \times {\omega _c}}&0
			\end{array}\right],\\
			{\kappa _l}&: = \frac{{n!}}{{(n + 1 - l)!(l - 1)!}},\quad l \in \left\{ {1, \ldots ,n} \right\}.
		\end{aligned}
	\end{equation}
	
	\subsection{Unified Estimation Framework for Dynamical Bounds}
	\label{sec.3b}
	In modern control theory,  methods based on Lyapunov functionals provide explicit time-domain criteria for optimizing stability margins. By constructing appropriate Lyapunov functionals and rescaling matrices,
	Theorems \ref{thm.1}--\ref{thm.2} in this subsection establishe a quantitative stability analysis paradigm for generic ADRC-based systems, with implications for broader control problems. The theorems will later be applied in Sec.~\ref{sec.3c} to the specific DLADRC system in this study.
	
	\subsubsection{Technical Theorem}
	
	The core analytical tool for establishing observation error bounds is the following theorem.
	
	\begin{Theorem} \label{thm.1}
		Let $\boldsymbol{P}$ be symmetric with  $m\left( \boldsymbol{P} \right)$ and $M\left( \boldsymbol{P}\right)$ representing its minimal and maximal eigenvalues. For integer $\nu  \geqslant 1$, define ${{\rm X}_\nu }: = C\left( {{\mathbb{R}_{ \geqslant 0}},{\mathbb{R}^\nu }} \right) \cap {C^1}\left( {{\mathbb{R}_{ > 0}},{\mathbb{R}^\nu }} \right)$. Let $\boldsymbol{Q}$ be a $D \times D$, $D \geqslant 1$ matrix satisfying, for some constant $\Omega  > 0$ and positive definite matrix $\boldsymbol{P}$,
		\begin{equation} \label{eq.11}
			{\boldsymbol{Q}^ \top }\boldsymbol{P} + \boldsymbol{PQ} =  - \Omega {\boldsymbol{I}_D}.
		\end{equation}
		Take $\boldsymbol{f} \in C\left( {{\mathbb{R}_ + },{\mathbb{R}^D}} \right)$ with ${\left\| \boldsymbol{f} \right\|_\infty }: = {\sup _{t \geqslant 0}}\left\| \boldsymbol{f} \right\| < \infty $. Then, for any $\boldsymbol{\eta} \in {{\rm X}_D}$ satisfying
		\begin{equation} \label{eq.12}
			\boldsymbol{\dot \eta} \left( t \right) = \boldsymbol{Q\eta} \left( t \right) + \boldsymbol{f}\left( t \right),\:t > 0,
		\end{equation}
		there holds
		\begin{equation} \label{eq.13}
			\left\| {\boldsymbol{\eta} \left( t \right)} \right\| \leqslant {c_1}{e^{ - \gamma t}}\left\| {\boldsymbol{\eta} \left( 0 \right)} \right\| + {c_2},\:t \geqslant 0,
		\end{equation}
		where ${c_1}: = \frac{M(\boldsymbol{P})}{m(\boldsymbol{P})}$, $c_2 = \frac{2M(\boldsymbol{P})^2 \left\| \boldsymbol{f} \right\|_\infty}{m(\boldsymbol{P})\Omega}$, and $\gamma = \frac{\Omega}{M(\boldsymbol{P})}$.
	\end{Theorem}
	
	\begin{Proof}
		Define the Lyapunov functional $V\left( t \right): = \left\langle {\boldsymbol{\eta} \left( t \right),\boldsymbol{P\eta} \left( t \right)} \right\rangle $ with $\left\langle \boldsymbol{v,w} \right\rangle  = {\boldsymbol{v}^ \top }\boldsymbol{w}$. By definition, it follows $V\left( t \right) \leqslant M\left( \boldsymbol P \right){\left\| {\boldsymbol \eta \left( t \right)} \right\|^2}$, $V\left( t \right) \geqslant m\left( \boldsymbol P \right){\left\| {\boldsymbol \eta \left( t \right)} \right\|^2}$. By \eqref{eq.11}--\eqref{eq.12}, the upper bound above for $V\left( t \right)$, and the Cauchy-Schwarz inequality, there holds $\frac{d}{dt}V(t)\leqslant  - \Omega {\left\| {\boldsymbol \eta \left( t \right)} \right\|^2} + 2M\left( \boldsymbol P \right){\left\| \boldsymbol f \right\|_\infty }\left\| {\boldsymbol \eta \left( t \right)} \right\|$. This, together with the upper bound $\frac{\Omega V(t)}{M(\boldsymbol{P})}\leqslant \Omega {\left\| {\boldsymbol \eta \left( t \right)} \right\|^2}$, implies
		\begin{equation} \label{eq.14}
			\left( {\frac{d}{{dt}} + \frac{\Omega }{{M\left( \boldsymbol P \right)}}} \right)V\left( t \right) \leqslant 2M(\boldsymbol P){\left\| \boldsymbol f \right\|_\infty }\left\| {\boldsymbol \eta \left( t \right)} \right\|.
		\end{equation}
		Multiplying both sides of \eqref{eq.14} by $\exp \left\{ {\frac{\Omega }{{M\left(\boldsymbol P \right)}}t} \right\}$, integrating over $\left( {0,t} \right)$, and using the lower and upper bound for $V\left( t \right)$ yield
		\begin{equation} \label{eq.15}
			\begin{aligned}
				m\left( \boldsymbol P \right)\exp \left\{ {\frac{\Omega }{{M\left(\boldsymbol P \right)}}t} \right\}&{\left\| {{\boldsymbol{\eta} _t}} \right\|^2}  \leqslant M\left( \boldsymbol P \right){\left\| {\boldsymbol \eta \left( 0 \right)} \right\|^2}   \\ 
				+ \frac{{2M{{\left(\boldsymbol P \right)}^2}{{\left\|\boldsymbol f \right\|}_\infty }}}{\Omega }
				&\times \left( {\exp \left\{ {\frac{\Omega }{{M\left(\boldsymbol P \right)}}t} \right\} - 1} \right)N\left( t \right), 
			\end{aligned} 
		\end{equation}
		where $N\left( t \right): = \mathop {\sup }\limits_{0 \leqslant \tau \leqslant t} \left\| {{\boldsymbol \eta _\tau }} \right\|$. Dividing both sides of \eqref{eq.15} by $m\left(\boldsymbol P \right)\exp \left\{ {\frac{\Omega }{{M\left(\boldsymbol P \right)}}t} \right\}N\left( t \right)$ gives the desired \eqref{eq.13}.
		\qed
	\end{Proof}
	\subsubsection{Unified Bounds for Cascade Structure}
	
	Based on Theorem \ref{thm.1}, we develop a unified explicit decay estimation framework for a general class of state space equations arising from ADRC-based systems with cascade ESOs, encompassing the DLADRC system in this study. By subtracting \eqref{eq.7} from \eqref{eq.4}, we obtain the differential equation governing the evolution of $\boldsymbol{\tilde z}_i=\boldsymbol{z}-\boldsymbol{ \hat z }_i$,
	
	\begin{equation} \label{eq.16}
		\left\{
		\begin{aligned}
			{\boldsymbol{\dot {\tilde z}}_1} &= (\boldsymbol A - {\boldsymbol l_1}{\boldsymbol c^ \top }){\boldsymbol{\tilde z}_1} + {\boldsymbol l_1}\rho  \\ 
			{\boldsymbol{\dot {\tilde z}}_i} &= (\boldsymbol A - {\boldsymbol l_i}{\boldsymbol c^ \top }){{\boldsymbol{\tilde z}}_i} + ({\boldsymbol l_i}{\boldsymbol c^ \top } - \boldsymbol b{\boldsymbol b^ \top }{\boldsymbol l_{i - 1}}{\boldsymbol c^ \top }){\boldsymbol{\tilde z}_{i - 1}} \\
			&+ \boldsymbol b{\boldsymbol b^ \top }{\boldsymbol l_1}\rho - \boldsymbol b{\boldsymbol b^ \top }\sum\limits_{k' = 1}^{i - 2} {\left( {{\boldsymbol l_{k'}}{\boldsymbol c^ \top } - {\boldsymbol l_{k' + 1}}{\boldsymbol c^ \top }} \right){\boldsymbol{\tilde z}_{k'}}} \\
			&\hspace{45mm} i \in \left\{ {2, \ldots ,p} \right\} 
		\end{aligned}  
		\right..
	\end{equation}
	The next theorem yields decay estimate for solutions to a general class of differential equations of the form \eqref{eq.16}. 
	\begin{Theorem} \label{thm.2}
		Let $i \geqslant 2$ and take $\rho, \:{h_k} \in C\left( {{\mathbb{R}_{ \geqslant 0}},\mathbb{R}} \right)$, with ${\left\| \rho  \right\|_\infty }, \:{\left\| {{h_k}} \right\|_\infty } < \infty $. Let $\boldsymbol{\tilde z} \in {{\rm X}_{n + 1}}$ solve
		\begin{equation} \label{eq.17}
			\boldsymbol{\dot{ \tilde z}} = (\boldsymbol A - {\boldsymbol l_i}{\boldsymbol c^ \top })\boldsymbol{\tilde z} + \boldsymbol{\bar f}, \:t > 0,
		\end{equation}
		where
		\begin{equation} \label{eq.18}
			\begin{aligned}
				\boldsymbol{\bar f} &:= \boldsymbol b{\boldsymbol b^ \top }{\boldsymbol l_1}\rho  + ({\boldsymbol l_i} - \boldsymbol b{\boldsymbol b^ \top }{\boldsymbol l_{i - 1}}){h_{i - 1}} \\
				&- \boldsymbol b{\boldsymbol b^ \top }\sum\limits_{k' = 1}^{i - 2} {({\boldsymbol l_{k'}} - {\boldsymbol l_{k' + 1}})} {h_{k'}}.
			\end{aligned}	
		\end{equation}
		Let $\boldsymbol \gamma \in \mathbb{R}^{n+1}$ with ${\gamma _m}$ as in \eqref{eq.6}. Write ${\operatorname{ls}_\infty }: = \lim {\sup _{t \to \infty }}$. Then there exist ${c_3}$, ${c_4}$, ${c_5}$ depending only on $n$ s.th. for all $t \geqslant 0$, and for each entry ${\tilde z^j}$, there holds
		\begin{equation} \label{eq.19}
			\begin{aligned}
				\left| {{{\tilde z}^j}\left( t \right)} \right| &\leqslant \omega _{oi}^{j - 1}{c_3}{e^{ - {c_4}{\omega _{oi}}t}}\left\| {\boldsymbol {\tilde z} \left( 0 \right)} \right\|\\
				&+ {c_5}\frac{1}{{\omega _{oi}^{n - j + 2}}} \{ \omega _{o1}^{n + 1}{\left\| \rho  \right\|_\infty }\\
				&+ \omega _{oi}^{n + 1}\left\| {\boldsymbol \gamma  - \boldsymbol b} \right\|{\left\| {{h_{i - 1}}} \right\|_\infty } \\
				&+ \sum\limits_{k = 1}^{i - 1} {\left( {\omega _{o\left( {k + 1} \right)}^{n + 1} - \omega _{ok}^{n + 1}} \right){{\left\| {{h_k}} \right\|}_\infty }} \},
			\end{aligned}	
		\end{equation}
		\begin{equation} \label{eq.20}
			\begin{aligned}
				{\operatorname{ls}_\infty \left| {{{\tilde z}^j}(t)} \right|} &\leqslant {c_5}\frac{1}{{\omega _{oi}^{n - j + 2}}}\{ \omega _{o1}^{n + 1}{\operatorname{ls}_\infty \left| \rho  \right|} \\
				&+ \omega _{oi}^{n + 1}\left\| {\boldsymbol \gamma  - \boldsymbol b} \right\|{\operatorname{ls}_\infty \left| {{h_{i - 1}}} \right|} \\
				&+ \sum\limits_{k = 1}^{i - 1} {\left( {\omega _{o\left( {k + 1} \right)}^{n + 1} - \omega _{ok}^{n + 1}} \right){\operatorname{ls}_\infty \left| {{h_k}} \right|}} \} .
			\end{aligned}	
		\end{equation}
	\end{Theorem}
	
	\begin{Proof} \label{proof.2}
		\enspace
		\par 
		\textbf{Step 1:} Let ${\boldsymbol L_i}: = \operatorname{diag} \left( {1, \ldots ,\omega _{oi}^n} \right)$ be the scaling matrix, thereby decoupling the ${\omega _{oi}}$'s from $\boldsymbol A - {\boldsymbol l_i}{\boldsymbol c^ \top }$ via the identity
		\begin{equation} \label{eq.21}
			\boldsymbol L_i^{ - 1}(\boldsymbol A - {\boldsymbol l_i}{\boldsymbol c^ \top }){\boldsymbol L_i} = {\omega _{oi}}\boldsymbol \Gamma 
		\end{equation}
		where ${\Gamma _{ij}} = - {\gamma _i}$ for $j = 1$, $= 1$ for $j = i + 1$, and $= 0$ elsewhere. $\boldsymbol \Gamma$ has  $ - 1$ as its only eigenvalue with multiplicity $n+1$, which guarantees the existence of a positive definite solution $\boldsymbol P$ to \eqref{eq.11} depending exclusively on $n$ (via $\boldsymbol \Gamma$) and $\Omega  = {\omega _{oi}}$\cite{lancaster1970}.
		
		\textbf{Step 2:} Set
		\begin{equation} \label{eq.22}
			\left( {\boldsymbol \eta, \: \boldsymbol Q, \: \boldsymbol f} \right): = (\boldsymbol L_i^{ - 1}\boldsymbol{\tilde z}, \: {\omega _{oi}}\boldsymbol \Gamma , \: \boldsymbol L_i^{ - 1}\boldsymbol{\bar f}).
		\end{equation}
		Through identity \eqref{eq.21}, the function $\boldsymbol \eta $ satisfies \eqref{eq.12}. Consequently, by Theorem \ref{thm.1}, there exist $c_3$, $c_4$, $c_5$ depending only on $n$ (via $\boldsymbol \Gamma$) s.th.
		\begin{equation} \label{eq.23}
			\left\| {\boldsymbol \eta \left( t \right)} \right\| \leqslant {c_3}{e^{ - {c_4}{\omega _{oi}}t}}\left\| {\boldsymbol \eta \left( 0 \right)} \right\| + {c_5}\omega _{oi}^{ - 1}{\left\| \boldsymbol f \right\|_\infty }.
		\end{equation}
		
		\textbf{Step 3:} To recover estimate for each entry of $\boldsymbol {\tilde z}$, recall that by definition \eqref{eq.22}, there holds $\left\| \boldsymbol \eta  \right\| \geqslant \left| {{\eta ^j}} \right| = \omega _{oi}^{ - j + 1}\left| {{{\tilde z}^j}} \right|$ (since ${\omega _{oi}} \geqslant 1$). This fact, together with estimate \eqref{eq.23}, yield
		\begin{equation} \label{eq.24}
			\left| {{{\tilde z}^j}\left( t \right)} \right| \leqslant {c_3}\omega _{oi}^{j - 1}{e^{ - {c_4}{\omega _{oi}}t}}\left\| {\boldsymbol{\tilde z}\left( 0 \right)} \right\| + {c_5}\omega _{oi}^{j - 2}{\left\| \boldsymbol f \right\|_\infty }.
		\end{equation}
		To bound ${\left\| \boldsymbol f \right\|_\infty }$, note by \eqref{eq.23}, there holds
		\begin{equation} \label{eq.25}
			\boldsymbol L_i^{ - 1}{\boldsymbol l_i} = {\omega _{oi}}\boldsymbol \gamma , \: \boldsymbol L_i^{ - 1}\boldsymbol b{\boldsymbol b^ \top }{\boldsymbol l_{i^\prime }} = \frac{{\omega _{oi^\prime }^{n + 1}}}{{\omega _{oi}^n}}\boldsymbol b,
		\end{equation}
		for any $i^\prime  \leqslant i$. Using definition \eqref{eq.18} and relations \eqref{eq.25} yields
		\begin{equation} \label{eq.26}
			\begin{aligned}
				\boldsymbol f &= \omega _{oi}^{ - n} \{ \omega _{oi}^{n + 1}\boldsymbol b\rho  + \omega _{oi}^{n + 1}\left( {\boldsymbol \gamma  - \boldsymbol b} \right){h_{i - 1}} \\
				&- \sum\limits_{k = 1}^{n - 1} {\left( {\omega _{ok}^{n + 1} - \omega _{o\left( {k + 1} \right)}^{n + 1}} \right)\boldsymbol b{h_k}}  \}.
			\end{aligned}	
		\end{equation}
		It follows that
		\begin{equation} \label{eq.27}
			\begin{aligned}
				{\left\| \boldsymbol f \right\|_\infty } &\leqslant \omega _{oi}^{ - n}\{ \omega _{oi}^{n + 1}{\left\| \rho  \right\|_\infty } + \omega _{oi}^{n + 1}\left\| {\boldsymbol \gamma - \boldsymbol b} \right\|{\left\| {{h_{i - 1}}} \right\|_\infty } \\
				& + \sum\limits_{k = 1}^{i - 1} {\omega _{o\left( {k + 1} \right)}^{n + 1} - \omega _{ok}^{n + 1}{{\left\| {{h_k}} \right\|}_\infty } \} } .
			\end{aligned}
		\end{equation}
		Inserting \eqref{eq.27} into \eqref{eq.24} gives \eqref{eq.19}.
		
		\textbf{Step 4:} To prove \eqref{eq.20}, take any $T > 0$ and write ${g_T}\left( t \right) := g\left( {t + T} \right)$. By definition, ${\boldsymbol{\tilde z}_T},\:T > 0$ satisfies \eqref{eq.17} with $\boldsymbol{\bar f}$ replaced by $\boldsymbol{\bar f}_T$. Proceeding as in Steps 1--3 above, we conclude that  ${\boldsymbol{\tilde z}_T}$ obeys \eqref{eq.24} with $\boldsymbol{f}$ replaced by $\boldsymbol{f}_T$. Since, moreover, ${\left\| {{\boldsymbol f_T}} \right\|_\infty } = \mathop {\sup }\limits_{t \geqslant T} \left\| \boldsymbol f \right\|$, it follows that for any $ \epsilon > 0$, there exists ${t_0} > 0$ s.th. $\mathop {\sup }\limits_{t \geqslant {t_0}} \left| {\tilde z_T^j \left( t \right)} \right| \leqslant {c_5}\omega _{oi}^{j - 2}\mathop {\sup }\limits_{t \geqslant T} \left\| {\boldsymbol f \left( t \right)} \right\| + \epsilon$. Since $T$ is arbitrary, sending $T \to \infty $ and using that $\mathop {\lim }\limits_{T \to \infty } \mathop {\sup }\limits_{t \geqslant {t_0}} \left| {{{\tilde z}^j}_T\left( t \right)} \right| = {\operatorname{ls}_\infty \left| {{{\tilde z}^j}(t)} \right|}$ for any ${t_0}$, yields ${\operatorname{ls}_\infty \left| {{{\tilde z}^j}(t)} \right|} \leqslant {c_5}\omega _{oi}^{j - 2}{\operatorname{ls}_\infty \left\| {f(t)} \right\|} + \epsilon$. Finally, \eqref{eq.20} follow by letting   $\epsilon \to {0_+ }$ and using \eqref{eq.26}.
		\qed
	\end{Proof}
	
	\subsection{Stability Analysis of the DLADRC Architecture}\label{sec.3c}
	
	This subsection establishes bounds for observation and control errors corresponding to $\boldsymbol{\tilde x}$,  $\boldsymbol{\tilde z}_p$, and $e$ in the state space equation \eqref{eq.5}, \eqref{eq.7}, and \eqref{eq.4}, where Assumption \ref{ass.1} ensures bounded ${\left\| {\dot F} \right\|_\infty }$ and ${\left\| \rho  \right\|_\infty }$, crucial for invoking Theorems \ref{thm.1} and \ref{thm.2} in our analysis.
	
	\subsubsection{Bounds for Inner Loop Observation}
	
	\begin{Theorem} \label{thm.3}
		Let $\boldsymbol{\tilde x} \in {{\rm X}_{n + 1}}$ solve $\boldsymbol{\dot {\tilde x}} = \left( {\boldsymbol A - \boldsymbol L{\boldsymbol c^ \top }} \right)\boldsymbol{\tilde x} + \boldsymbol b\dot F, t > 0$. Then, for ${c_3}$, ${c_4}$, ${c_5}$ as in Theorem \ref{thm.2} and $\forall t \geqslant 0$,
		\begin{equation}
			\left| {{{\tilde x}^j}\left( t \right)} \right| \leqslant {c_3}\omega _{{\text{in}}}^{j - 1}{e^{ - {c_4}{\omega _{{\text{in}}}}t}}\left\| {\boldsymbol{\tilde x}\left( 0 \right)} \right\| + \frac{{{c_5}}}{{\omega _{{\text{in}}}^{n - j + 2}}}{\left\| {\dot F} \right\|_\infty },
		\end{equation}
		\begin{equation}
			{\operatorname{ls}_\infty \left| {{{\tilde x}^j}\left( t \right)} \right|} \leqslant \frac{{{c_5}}}{{\omega _{{\text{in}}}^{n - j + 2}}}{\left\| {\dot F} \right\|_\infty }.
		\end{equation}
	\end{Theorem}
	
	\begin{Proof}
		Let $\boldsymbol \Lambda : = \operatorname{diag} \left( {1, \ldots ,\omega _{{\text{in}}}^n} \right)$. We have ${\boldsymbol L^{ - 1}}\left( {\boldsymbol A - \boldsymbol L{\boldsymbol c^ \top }} \right)\boldsymbol L = {\omega _{{\text{in }}}}\boldsymbol \Gamma $, where $\boldsymbol \Gamma$ is as in \eqref{eq.21}. Set $\left( {\boldsymbol \eta, \: \boldsymbol Q, \: \boldsymbol f} \right): = \left( {{\boldsymbol \Lambda ^{ - 1}}\boldsymbol {\tilde x}, \: {\omega _{{\text{in }}}}\boldsymbol \Gamma , \: {\boldsymbol \Lambda ^{ - 1}}\boldsymbol b\dot F} \right)$. Then it follows from Theorem \ref{thm.1} that $\left\| {\boldsymbol \eta \left( t \right)} \right\| \leqslant {c_3}{e^{ - {c_4}{\omega _{{\text{in}}}}t}}\left\| {\boldsymbol \eta \left( 0 \right)} \right\| + {c_5}\omega _{{\text{in}}}^{ - 1}{\left\| \boldsymbol f \right\|_\infty }$. Since ${\boldsymbol \Lambda ^{ - 1}}\boldsymbol b = \omega _{{\text{in}}}^{ - n}\boldsymbol b$ and therefore $\boldsymbol f = \omega _{{\text{in}}}^{ - n}\boldsymbol b\dot F$, proceeding exactly as in Step 4 in the proof of Theorem \ref*{thm.2}  yields the desired estimates.
		\qed
	\end{Proof}
	
	\subsubsection{Bounds for Outer Loop Observation}
	
	\begin{Theorem} \label{thm.4}
		Let ${\boldsymbol{\tilde z}_1}, \ldots ,{\boldsymbol{\tilde z}_p} \in {{\rm X}_{n + 1}}$ solve \eqref{eq.16}. Let ${\omega _{oi}} = {\alpha ^{i - 1}}{\omega _{o1}}$ for some fixed $\alpha > 1$. Then there exist ${c_8},\:{c_9} > 0$ depending only on $n$, $p$, $\alpha$ s.th. for all $t \geqslant 0$,
		\begin{equation} \label{eq.30}
			\begin{aligned}
				\left| {\tilde z_p^j\left( t \right)} \right| &\leqslant {c_9}\omega _{op}^{j - 1} \{ {{e^{ - {c_4}{\omega _{op}}t}}\left\| {{\boldsymbol {\tilde z}_p}\left( 0 \right)} \right\| }\\
				&{+ \sum\limits_{i = 1}^{p - 1} {\left\| {{\boldsymbol {\tilde z}_i}\left( 0 \right)} \right\|}  + {{\left\| \rho  \right\|}_\infty }} \},
			\end{aligned}
		\end{equation}
		\begin{equation} \label{eq.31}
			{\operatorname{ls} _\infty \left| {\tilde z_p^j(t)} \right|} \leqslant {c_8}\omega _{op}^{j - 1}{\left\| \rho  \right\|_\infty }.
		\end{equation}
	\end{Theorem}
	
	\begin{Proof}
		We start with ${\boldsymbol {\tilde z}_1}$. Using the relation $\boldsymbol L_1^{ - 1}{\boldsymbol l_1}\rho  = {\omega _{o1}}\boldsymbol \gamma \rho$ instead of \eqref{eq.25} and proceeding as in Step 4 in the proof of Theorem \ref{proof.2} lead to ${\operatorname{ls} _\infty {\left| {\tilde z_1^j(t)} \right|}} \leqslant {c_6}\omega _{o1}^{j - 1}{\left\| \rho  \right\|_\infty },\: {c_6}: = {c_5}\left\| \boldsymbol \gamma  \right\|$. Now we bound ${\boldsymbol{\tilde z}_i},\:i = 2, \ldots ,p$ by induction on $p$. For the base case $p = 2$, consider the equation for ${\boldsymbol{\tilde z}_2}$, which is of the form \eqref{eq.17} upon choosing ${h_1} = z_1^1$. By Theorem \ref{thm.2} and the estimate for ${\boldsymbol{\tilde z}_1}$, we find ${\operatorname{ls}_\infty \left| {\tilde z_2^j} \right|} \leqslant {c_7}\omega _{o2}^{j - 1}{\left\| \rho  \right\|_\infty }$, where ${c_7}: = {c_5}\left( {{{\omega _{o1}^{n + 1}} \mathord{\left/ {\vphantom {{\omega _{o1}^{n + 1}} {\omega _{o2}^{n + 1}}}} \right. \kern-\nulldelimiterspace} {\omega _{o2}^{n + 1}}} + {c_6}\left\| {\boldsymbol \gamma  - \boldsymbol b} \right\|} \right)$. This proves \eqref{eq.31} for  $p = 2$.
		
		Now, for the induction hypothesis, assume
		\begin{equation} \label{eq.32}
			{\operatorname{ls} _\infty }\left| {\tilde z_k^j} \right| \leqslant {C_k}\omega _{ok}^{j - 1}{\left\| \rho  \right\|_\infty },\:k = 1, \ldots ,p - 1
		\end{equation}
		We will prove the same for $k = p$, which is exactly (31).
		
		Consider the equation for ${\boldsymbol{\tilde z}_p}$, which is of the form \eqref{eq.17} with ${h_k} = z_k^1,\:k = 1, \ldots ,p - 1$. Applying Theorem \ref{thm.2}, together with the induction hypothesis \eqref{eq.32}, yields ${\operatorname{ls} _\infty \left| {\tilde z_p^j} \right| } \leqslant {c_8}\omega _{op}^{j - 1}{\left\| \rho  \right\|_\infty }$, where ${c_8}: = {c_5} \left\{ {{\omega _{oi}^{n + 1}} \mathord{\left/ {\vphantom {{\omega _{oi}^{n + 1}} {\omega _{op}^{n + 1}}}} \right. \kern-\nulldelimiterspace} {\omega _{op}^{n + 1}}} + {C_{p - 1}}\left\| {\boldsymbol \gamma  - \boldsymbol b} \right\| + \sum\limits_{i = 1}^{p - 1} {{C_i}} \left( {\omega _{o(i + 1)}^{n + 1} - \omega _{oi}^{n + 1}} \right)\right\} $ depends only on $n$, $\alpha $, and ${C_1}, \ldots ,{C_{p - 1}}$. This completes the induction step and \eqref{eq.31} is proved.                                                                                      
		\qed
	\end{Proof}
	
	\subsubsection{Bounds for Control Error}
	
	\begin{Theorem}
		\label{thm.5}
		Define $\boldsymbol \epsilon: = {\left[ {\begin{array}{*{20}{c}} e&{\dot e}& \cdots &{{e^{(n - 1)}}} \end{array}} \right]^ \top }$ satisfying:
		\begin{equation} \label{eq.33}
			{\dot {\epsilon^q}} = {\boldsymbol c^ \top }{\boldsymbol A^q}\boldsymbol z = {z^{q + 1}} = {\epsilon^{q + 1}}, \:q = 1, \ldots ,n - 1,
		\end{equation}
		\begin{equation} \label{eq.34}
			{\dot {\epsilon^n}} = - \omega _{\text{c}}^n{\epsilon^1} - \sum\limits_{q = 1}^{n - 1} {K_{{\text{out }}}^{q + 1}} {\epsilon^{q + 1}} + {\boldsymbol K_{{\text{out }}}}{\boldsymbol{\tilde z}_p} + \omega _{\text{c}}^n\rho .
		\end{equation}
		Then there exist ${c_{10}},\:{c_{11}} > 0$ depending only on $n$, $p$, $\alpha $ s.th. for all $t \geqslant 0$,
		\begin{equation} \label{eq.35}
			\begin{aligned}
				\left| {{\epsilon^q}\left( t \right)} \right| &\leqslant {c_{11}}\omega _{\text{c}}^{q - 1} \{ {{e^{ - {c_4}{\omega _{\text{c}}}t}}\left\| \boldsymbol{\epsilon} {\left( 0 \right)} \right\|}\\
				& {+ \omega _{op}^n\sum\limits_{i = 1}^p {\left\| {{\boldsymbol{\tilde z}_i}\left( 0 \right)} \right\|} + \omega _{op}^n{{\left\| \rho  \right\|}_\infty }} \},
			\end{aligned}	
		\end{equation}
		\begin{equation} \label{eq.36}
			{\operatorname{ls}_\infty \left| {{\epsilon^q}(t)} \right| } \leqslant {c_{10}}\omega _{op}^n\omega _{\text{c}}^{q - 1}{\left\| \rho  \right\|_\infty }.
		\end{equation}
	\end{Theorem}
	
	\begin{Proof}
		Definitions of $\boldsymbol A$, $\boldsymbol c$, and $\boldsymbol d$ in \eqref{eq.3}, imply the relation ${\boldsymbol c^ \top }{\boldsymbol A^{p - 1}}\boldsymbol d = 0$ for $p = 1, \ldots ,n - 1$. Iteratively using this relation together with \eqref{eq.4} yields
		\begin{equation}
			{\dot {\epsilon^p}} = {e^{\left( p \right)}} = {\boldsymbol c^ \top }{\boldsymbol z^{\left( p \right)}} = {\boldsymbol c^ \top }{\boldsymbol A^{p - 1}}\left( {\boldsymbol {Az} - \boldsymbol d{u_{{\text{out}}}}} \right) = {\boldsymbol c^ \top }{\boldsymbol A^p}\boldsymbol z.
		\end{equation}
		Noting that ${\boldsymbol c^ \top }{\boldsymbol A^p}\boldsymbol z = {z^{p + 1}}$, we conclude \eqref{eq.33} from here.
		
		Next, we prove \eqref{eq.34}. Using \eqref{eq.33} with $p = n - 1$ and the relations ${\boldsymbol c^ \top }{\boldsymbol A^n}\boldsymbol d = 1$, ${\boldsymbol c^ \top }{\boldsymbol A^n}\boldsymbol z = {z^{n + 1}} = 0$ yields
		\begin{equation} \label{eq.38}
			{\dot{\epsilon^n}} = {\boldsymbol c^ \top }{\boldsymbol A^{n - 1}}\left( {\boldsymbol{Az} - \boldsymbol d{u_{{\text{out}}}}} \right) = \omega _{\text{c}}^n{y_{{\text{out}}}} + {\boldsymbol K_{{\text{out}}}}{\boldsymbol{\hat z}_p}.
		\end{equation}
		By the definition of $\boldsymbol \epsilon$ and the relation $\boldsymbol{\hat z} = \boldsymbol z - \boldsymbol{\tilde z}_p$, there holds
		\begin{equation} \label{eq.39}
			\begin{aligned}
				\omega _{\text{c}}^n{y_{{\text{out}}}} + {\boldsymbol K_{{\text{out}}}}{\boldsymbol{\hat z}_p} &= \omega _{\text{c}}^n{\epsilon^1} + \sum\limits_{q = 1}^{n - 1} {K_{{\text{out}}}^{q + 1}} {\epsilon^{q + 1}} \\
				&- {\boldsymbol K_{{\text{out}}}}{\boldsymbol{\tilde z}_p} + \omega _{\text{c}}^n\rho .
			\end{aligned}
		\end{equation}
		Inserting \eqref{eq.39} back to \eqref{eq.38} yields \eqref{eq.34}.
		
		To prove \eqref{eq.35}--\eqref{eq.36}, introduce the scaling matrix $\boldsymbol \Lambda  = \operatorname{diag} \left( {1, \ldots ,\omega _{\text{c}}^{n - 1}} \right)$ and set
		\begin{equation}
			\left( {\boldsymbol \eta, \: \boldsymbol Q, \: \boldsymbol f} \right): = \left( {{\boldsymbol \Lambda ^{ - 1}\boldsymbol \epsilon}, \: {\omega _{\text{c}}}\boldsymbol \Xi , \: {\boldsymbol \Lambda ^{ - 1}}\boldsymbol {\bar b} \left( {{\boldsymbol K_{{\text{out}}}}{\boldsymbol {\tilde z}_p} - \omega _{\text{c}}^n\rho } \right)} \right),
		\end{equation}
		where $\boldsymbol{\bar b}: = \left[ {\begin{array}{*{20}{l}} 0& \cdots &0&1 \end{array}} \right]_n^ \top $, ${\Xi _{ij}} =  - {\kappa _i}$ for $i = n$, $ = - 1$ for $i = j - 1$, and $ = 0$ elsewhere. Clearly, the matrix $\boldsymbol \Xi$ depends only on $n$ and the triple $\left( {\boldsymbol \eta ,\:\boldsymbol Q,\:\boldsymbol f} \right)$ satisfies \eqref{eq.12}. Hence, it follows from Theorem \ref{thm.1} that $\left\| {\boldsymbol \eta \left( t \right)} \right\| \leqslant {c_3}{e^{ - {c_4}{\omega _{\text{c}}}t}}\left\| {\boldsymbol \eta \left( 0 \right)} \right\| + {c_5}\omega _{\text{c}}^{ - 1}{\left\| \boldsymbol f \right\|_\infty }$. The identity ${\boldsymbol \Lambda ^{ - 1}}\boldsymbol{\bar b} = \omega _{\text{c}}^{ - (n - 1)}\boldsymbol{\bar b}$, the bound $\left\| {{\boldsymbol K_{{\text{out}}}}} \right\| \leqslant \omega _{\text{c}}^n$, and the estimates for $\boldsymbol {\tilde z}_p$ derived in Theorem \ref{thm.4} together yield
		\begin{equation} \label{eq.41}
			\left\| {\boldsymbol f\left( t \right)} \right\| \leqslant {\omega _{\text{c}}}\omega _{op}^n\left\| {{\boldsymbol{\tilde z}_p}\left( t \right)} \right\| + {\omega _{\text{c}}}{\left\| \rho  \right\|_\infty },\:t \geqslant 0,
		\end{equation}
		\begin{equation} \label{eq.42}
			{\operatorname{ls}_\infty \left\| {f\left( t \right)} \right\|} \leqslant {c_{10}}{\omega _{\text{c}}}\omega _{op}^n{\left\| \rho  \right\|_\infty }.
		\end{equation}
		Here, ${c_{10}} > 0$ depends only on $n$, $p$, $\alpha$ (through $c_8$, $c_9$ in \eqref{eq.30}, \eqref{eq.31} and   ${\kappa _1}, \ldots ,{\kappa _n}$). From \eqref{eq.41}--\eqref{eq.42} we proceed exactly as in Step 4 in the proof of Theorem \ref{proof.2} to conclude the desired estimates in \eqref{eq.35}--\eqref{eq.36}.
		\qed
	\end{Proof}
	
	\begin{figure}[!t]
		\centering
		\subfloat[(a)]
		{		
			\label{fig.4a}
			\begin{minipage}[h]{0.45\columnwidth}
				\includegraphics[width=\textwidth]{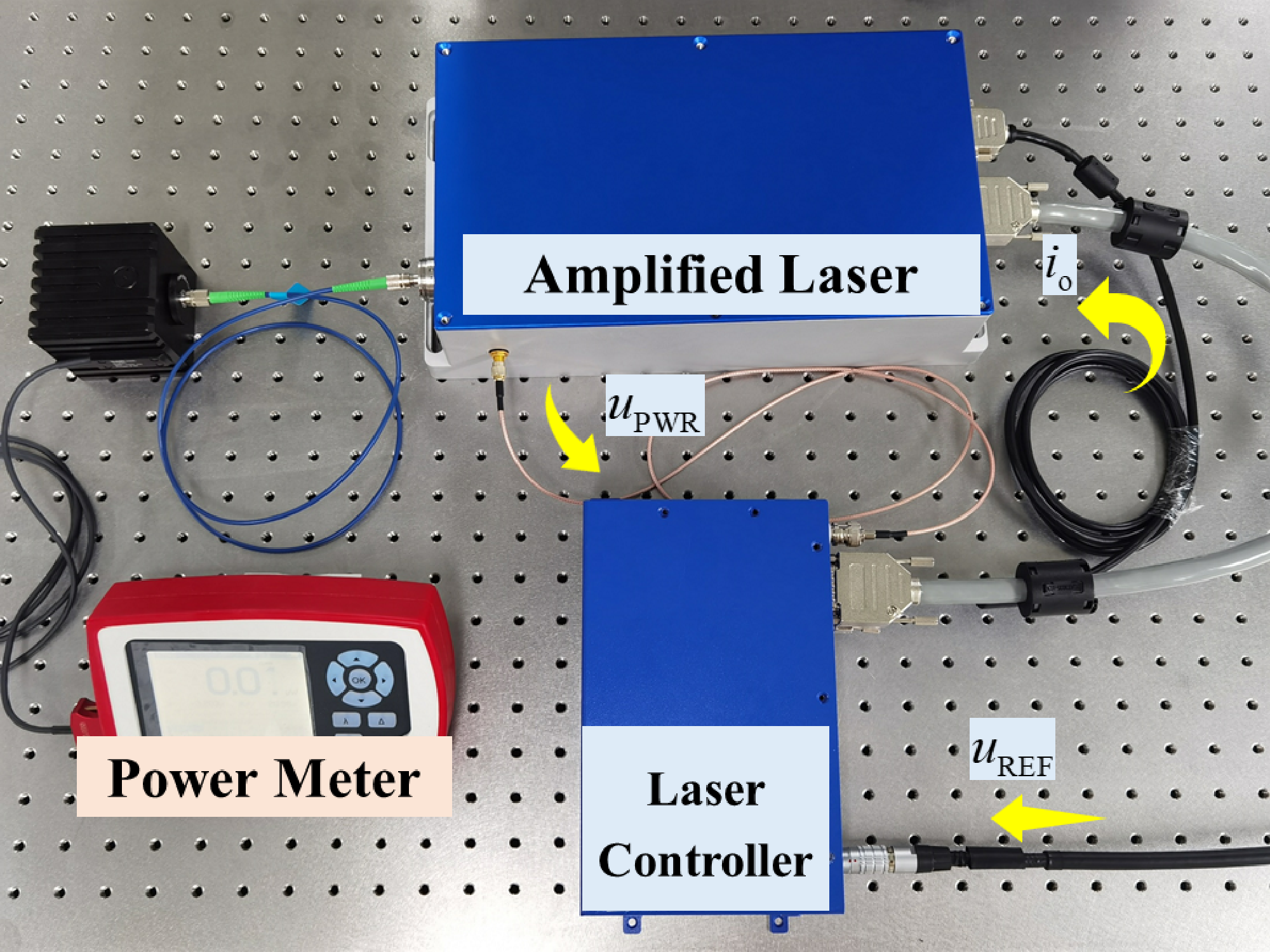}
			\end{minipage}
		}
		\subfloat[(b)]
		{		
			\label{fig.4b}
			\begin{minipage}[h]{0.45\columnwidth}
				\includegraphics[width=\textwidth]{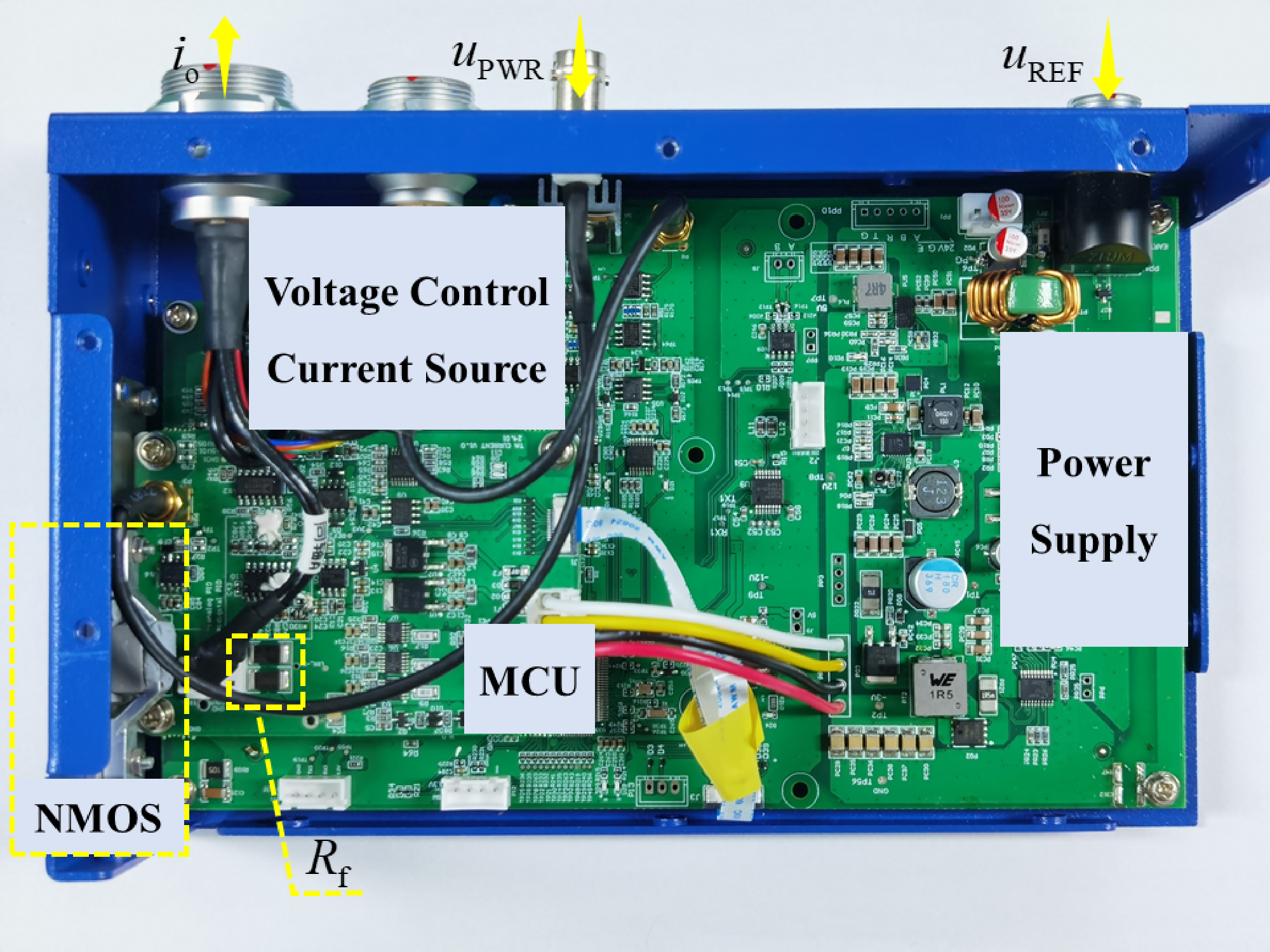}
			\end{minipage}
		}
		\caption{(a) Picture of the compact AL power stabilization experimental platform. (b) The internal view of the AL controller. MCU: Microcontroller Unit.}
		\label{fig.4}
	\end{figure}
	\begin{table}[!t]
		\renewcommand{\arraystretch}{1.3}
		\caption{Physical Parameters of Components in the Compact AL Power Stabilization System}
		\centering
		\label{table_1}
		\resizebox{\columnwidth}{!}{
			\begin{tabular}{l p{37mm} l}
				\hline\hline 
				\multicolumn{1}{c}{\textbf{Symbol}} & \multicolumn{1}{c}{\textbf{Parameters}} & \multicolumn{1}{c}{\pbox{20cm}{\textbf{Value}}}  \\
				\hline
				$ {\phi _{{\text{AL}}}} $  & AL conversion coefficient & $ 1.5 \times {10^{ - 3}}\:{{\mathrm{V}} \mathord{\left/ {\vphantom {{\mathrm{V}} {\mathrm{A}}}} \right. \kern-\nulldelimiterspace} {\mathrm{A}}} $ \\
				
				$ {\mu _{{\text{OA}}}} $ &Open-loop gain of the operational amplifier &  $ 2 \times {10^5}\: {{\mathrm{V}} \mathord{\left/{\vphantom {{\mathrm{V}} {\mathrm{V}}}} \right.\kern-\nulldelimiterspace}{\mathrm{V}}} $\\ 
				
				$ {f_{{\text{OAL}}}} $ & Frequency of the operational amplifier low pole & $ 1\:{\mathrm{Hz}} $ \\
				
				$ {f_{{\text{OAH}}}} $ & 
				Frequency of the operational amplifier high pole & $ 500\:{\mathrm{kHz}} $ \\ 
				
				$ {g_{\text{m}}} $ & Transconductance of MOSFET  & $ 0.2\:{\mathrm{A} \mathord{\left/ {\vphantom {{\mathrm{A}} {\mathrm{V}}}} \right. \kern-\nulldelimiterspace} {\mathrm{V}}} $ \\ 
				
				$ {C'_{{\text{gs}}}} $ & Total capacitance between the gate and source of MOSFET & $ 2.2\:\mathrm{\mu F} $ \\
				
				$ {R_{\text{c}}} $ & Compensation resistor & $ 200\:\mathrm{k\Omega}  $ \\
				
				$ {R_{\text{f}}} $  & Feedback resistor & $ 1\:\mathrm{m\Omega}$ \\
				
				\hline\hline
			\end{tabular}
		}
	\end{table}
	
	\section{Hardware Experiment}
	\label{sec.4}
	\subsection{Experimental Setup}
	\label{sec.4a}
	An experimental platform is developed to validate the efficacy of the proposed compact AL power stabilization approach based on DLADRC, as illustrated in Fig. \ref{fig.4}\subref{fig.4a}. Fig. \ref{fig.4}\subref{fig.4b} presents a detailed view of the laser control board implementation. As indicated by the signal flow arrows, the system forms a closed-loop control structure consistent with the schematic diagram in Fig. \ref{fig.1}. According to the device datasheets, the physical parameters are determined for providing numerical guidance, as summarized in Table \ref{table_1}. The following experiments are conducted to investigate the theoretical advantage and practical benefits of the proposed strategy.
	
	The system operates through the following stages.
	\subsubsection{Target Power Setting}
	The host computer transmits the target power value to the AL controller, configured according to the operational requirements of the OPM arrays. Power settings are specifically customized for each of the three experimental configurations with respective objectives.
	\subsubsection{Optical Feedback Acquisition}
	The probe beam is detected via a calibrated silicon photodetector (Thorlabs SM05PD3A) with subsequent digitization by a 16-bit analog-to-digital converter operating at a sampling rate of $250\:\mathrm{kHz}$. 
	\subsubsection{Control Processing}
	Following execution of the control algorithm, the MCU adjusts the AL's driving current via a high-precision digital-to-analog converter. 
	Standard ADRC (SADRC) and DLADRC serve as the baseline controllers for comparison, with parameter configuration in Table \ref{table_2}.
	
	\begin{figure}[!t]
		\centering
		\subfloat[(a)]
		{		
			\label{fig.5a}
			\begin{minipage}[h]{0.45\columnwidth}
				\includegraphics[height=\textwidth]{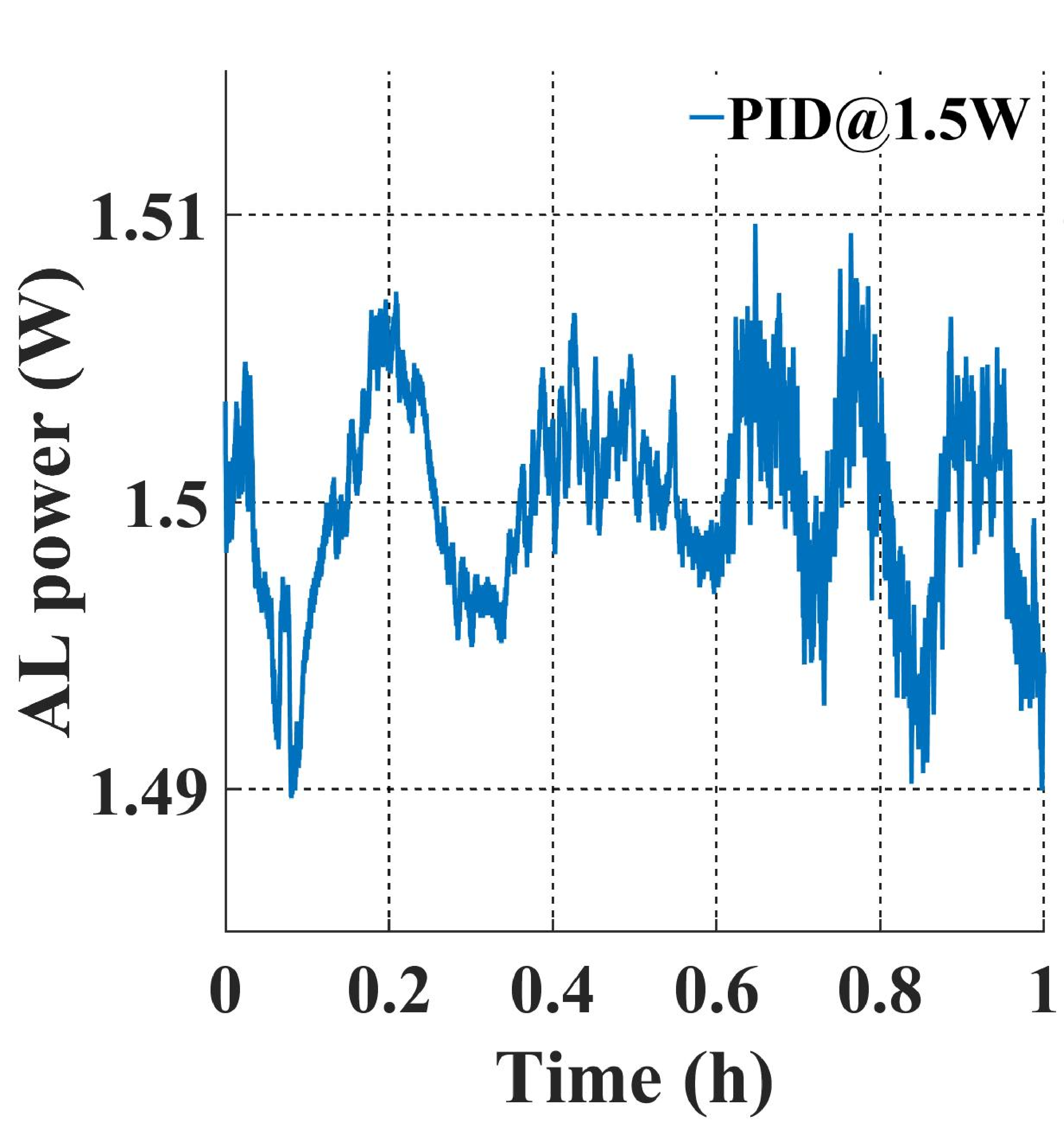}
			\end{minipage}
		}
		\subfloat[(b)]
		{		
			\label{fig.5b}
			\begin{minipage}[h]{0.45\columnwidth}
				\includegraphics[height=\textwidth]{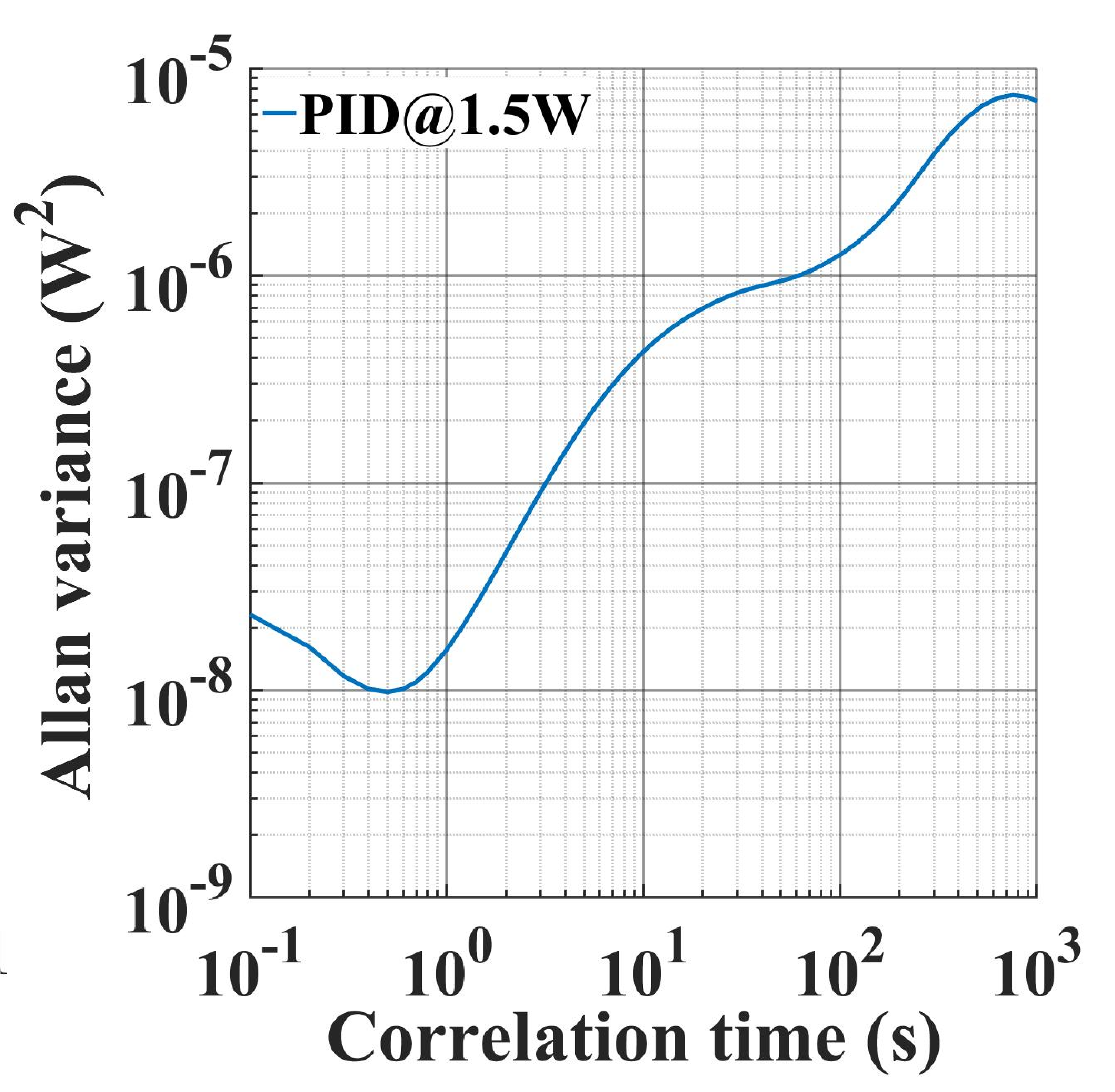}
			\end{minipage}
		}
		\caption{Performance with PID strategy. (a) 1-hour AL power fluctuation. (b) Allan variance of data in (a).}
		\label{fig.5}
	\end{figure}
	
	\subsubsection{Performance Verification}
	The AL power stability is measured using an integrating sphere power meter (Thorlabs S142C). We quantitatively define the stability metrics by Allan variance analysis across three timescales: short-term ($10^{-1}\:\mathrm{s}$ --$10^1\:\mathrm{s}$), mid-term ($10^1\:\mathrm{s}$--$10^2\:\mathrm{s}$), and long-term ($10^2\:\mathrm{s}$--$10^3\:\mathrm{s}$).
	
	To ensure a fair comparison between the SADRC and DLADRC strategies, we configure both with identical upper limits for the ESOs' bandwidth $\omega_{op}$ and the control bandwidth $\omega_{\rm{c}}$ (see Table \ref{table_2}). In the DLADRC, the observation range of outer loop is expanded compared to that of the inner loop to account for the photodetector measurement noise $\rho$, which predominantly occupies higher frequency bands relative to external disturbances and model parameter variations. Both controllers are evaluated under the same experimental setups, including hardware setup and parameter configurations.
	
	Besides, before employing ADRC-based approaches, the conventional proportional-integral-derivative (PID) control strategy demonstrates poor performance in the compact AL power stabilization system, as shown in Fig. \ref{fig.5}. The results reveal significant jitter and long-period oscillations (Fig. \ref{fig.5}\subref{fig.5a}), with corresponding notable degradation in stability evidenced by increased Allan variance (Fig. \ref{fig.5}\subref{fig.5b}).
	
	\subsection{Results and Discussion}
	\label{sec.4b}
	\subsubsection{Experiment E1 --- Verification of the Quantitative Control Estimate Predicted by Theory}
	
	To validate the theoretical control error estimate $\epsilon^1$ derived in \eqref{eq.35}, two opposing step signals are applied, i.e., a positive step from $1\:\mathrm{W}$ to $1.5\:\mathrm{W}$ and a negative step from $2\:\mathrm{W}$ to $1.5\:\mathrm{W}$. Furthermore, the transient response is continuously monitored throughout the 10-minute duration following each step input, providing a complete characterization of the convergence dynamics.
	\begin{figure}[!t]
		\centering
		\subfloat[(a)]
		{		
			\label{fig.6a}
			\begin{minipage}[h]{0.73\columnwidth}
				\includegraphics[width=\textwidth]{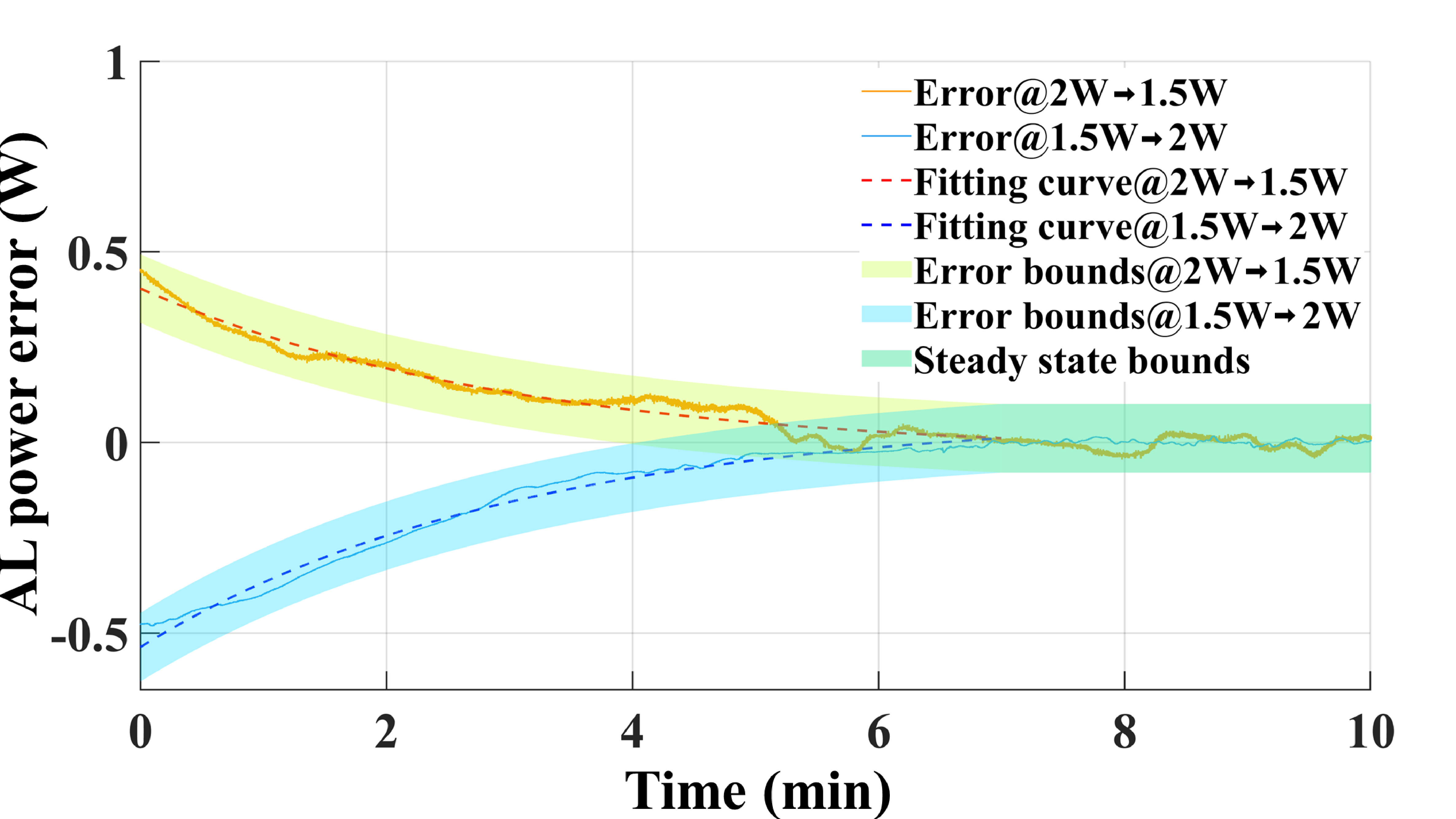}
			\end{minipage}
		}\\
		\subfloat[(b)]
		{		
			\label{fig.6b}
			\begin{minipage}[h]{0.73\columnwidth}
				\includegraphics[width=\textwidth]{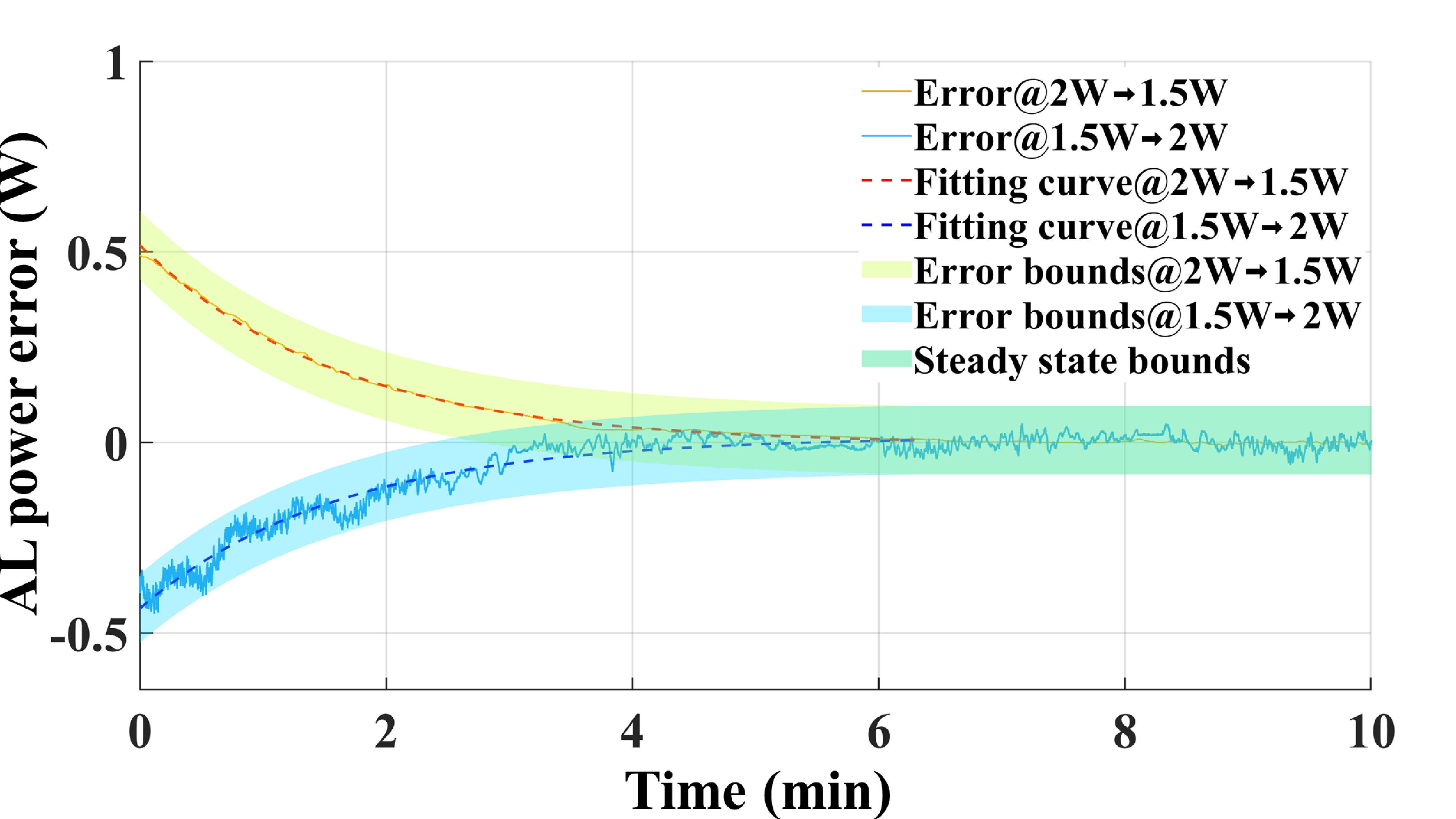}
			\end{minipage}
		}
		\caption{Results of E1. (a) ${\omega _{\text{c}}} = 150$. (b) ${\omega _{\text{c}}} = 300$.}
		\label{fig.6}
	\end{figure}
	
	\begin{table}[!t]
		\centering
		\caption{Bandwidth Configuration in Three Experiments}
		\renewcommand{\arraystretch}{1.3}
		\resizebox{\columnwidth}{!}{
			\begin{tabular}{ccccc}
				\hline\hline
				&\multirow{2}{*}{\textbf{Strategy}} &\multicolumn{3}{c}{\textbf{Parameters}}\\
				& & $\omega_{\rm{c}}$ & $\omega_{\rm{in}}$ & $\omega_{o1}$ \\
				\hline
				E1 & DLADRC ($p=3$) & $150$, $300$ & $800$ & $150$ \\
				
				\multirow{2}{*}{E2} 
				& SADRC & $300$ & --- & $1000$ \\
				& DLADRC ($p=1$) & $300$ & $800$ & $1000$ \\
				
				\multirow{2}{*}{E3} 
				& DLADRC ($p=2$) & $300$ & $800$ & $300$, $400$, $500$ \\
				& DLADRC ($p=3$) & $300$ & $800$ & $150$, $200$, $250$ \\
				\hline\hline
			\end{tabular}
		}
		\label{table_2}
	\end{table}
	
	Fig. \ref{fig.6} presents the results of E1, demonstrating strong agreement with theoretical predictions. For both crossover tested frequencies of ${\omega _{\text{c}}} = 150$ and ${\omega _{\text{c}}} = 300$, exponential fittings of the four datasets produce R-squared values exceeding 95\%, confirming the exponential decay behavior of the control error as predicted by \eqref{eq.35}. Moreover, with differences consistently below $0.3\%$, the decay coefficients exhibit remarkable consistency, measuring $0.335$ and $0.336$ when ${\omega _{\text{c}}} = 150$ and $0.617$ and $0.616$ when ${\omega _{\text{c}}} = 300$.  
	
	The near-identical decay rates across different step inputs support the theoretical conclusion derived in \eqref{eq.35} that ${\omega _{\text{c}}}$ predominantly governs the convergence process. This dominance is further evidenced by comparing the convergence periods required for the error to stabilize within $\pm 5\%$ of the step amplitude, as shown in Fig. \ref{fig.6}\subref{fig.6a} and \subref{fig.6b}. Doubling ${\omega _{\text{c}}}$ from $150$ to $300$ reduces the convergence time by over $25\%$, reinforcing the critical role of the controller bandwidth ${\omega _{\text{c}}}$ in determining the response dynamics of DLADRC system.
	\begin{figure}[!t]
		\centering
		\subfloat[(a)]
		{		
			\label{fig.7a}
			\begin{minipage}[h]{0.4\columnwidth}
				\includegraphics[width=\textwidth]{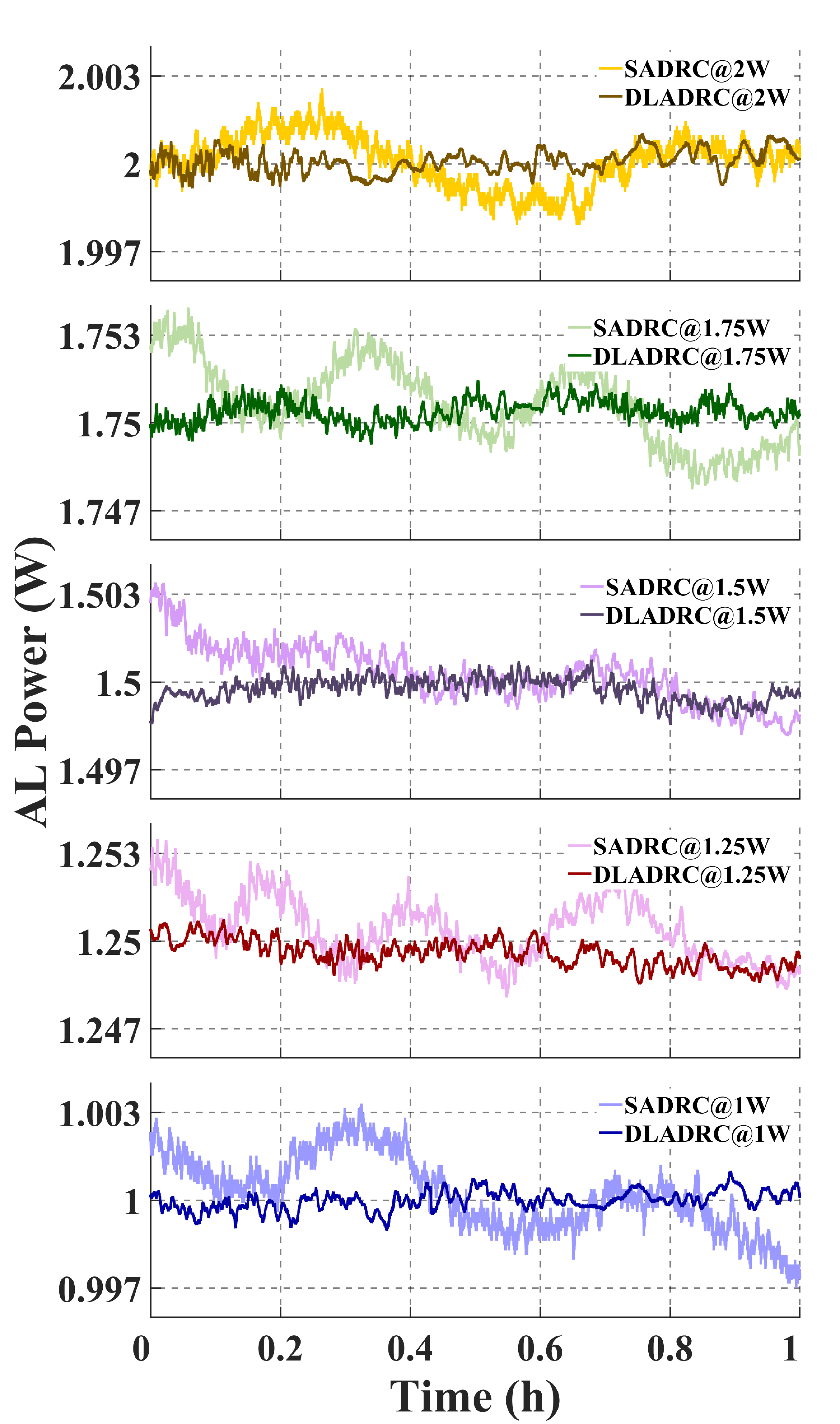}
			\end{minipage}
		}
		\subfloat[(b)]
		{		
			\label{fig.7b}
			\begin{minipage}[h]{0.55\columnwidth}
				\includegraphics[width=\textwidth]{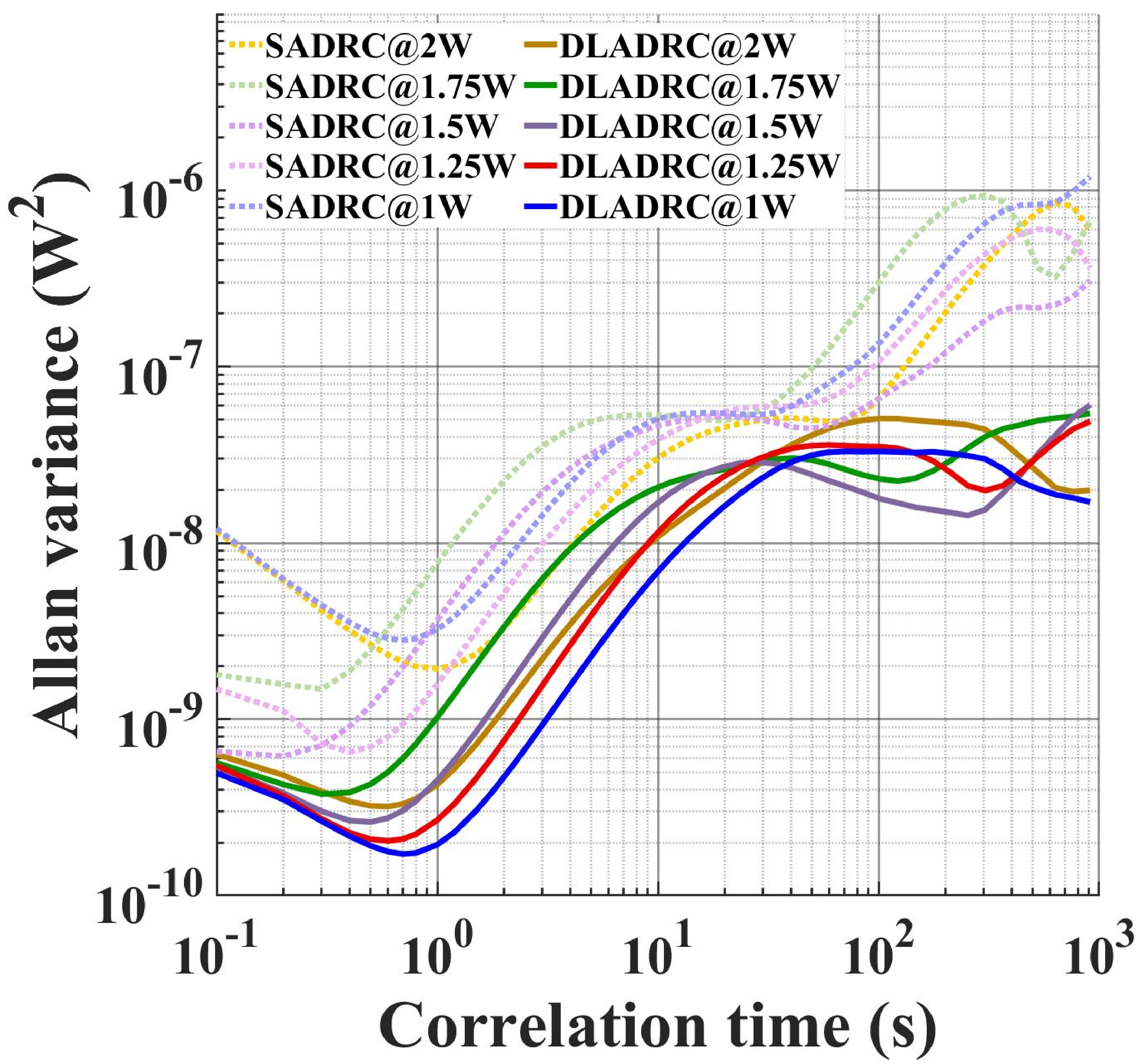}
			\end{minipage}
		}
		\caption{Results of E2. (a) Time domain. (b) Allan variance.}
		\label{fig.7}
	\end{figure}
	
	\begin{table}[!t]
		\caption{Comparison of AL Power Instability in E2}
		\centering
		\label{table_3}
		\renewcommand{\arraystretch}{1.3}
		\resizebox{\columnwidth}{!}{
			\begin{tabular}{cccccc}
				\hline\hline 
				\multirow{2}{*}{\textbf{Instability}} &\multicolumn{5}{c}{\textbf{Working points}}\\
				& $1\:\mathrm W$ & $1.25\:\mathrm W$ & $1.5\:\mathrm W$ & $1.75\:\mathrm W$ &$2\:\mathrm W$ \\
				\hline
				
				SADRC & $0.63\%$ & $0.43\%$ & $0.35\%$ & $0.35\%$ & $0.23\%$ \\
				
				DLADRC & $0.21\%$ & $0.18\%$ & $0.15\%$ & $0.13\%$ & $0.09\%$ \\
				
				Improvement & $66.7\%$ & $58.1\%$ & $57.1\%$ & $62.9\%$ & $60.9\%$ \\
				\hline\hline
			\end{tabular}
		}
	\end{table}
	\subsubsection{Experiment E2 --- Verification of Robustness Enhancement from the Supplemental Inner Loop}
	In contrast to E1, E2 focuses on the steady-state performance by evaluating the compact AL stabilization system at five distinct power levels spaced at $0.25\:\mathrm{W}$ intervals across the full $1\:\mathrm{W}$ to $2\:\mathrm{W}$ operational range of the AL. The assessment combines 1-hour power fluctuation measurements with Allan variance analysis to comprehensively characterize stability. 
	
	Fig. \ref{fig.7} presents key findings, highlighting the advantages of the inner-loop ESO's integration when the outer loop level set to $p=1$. The comparison in Fig. \ref{fig.7}\subref{fig.7a} reveals significant limitations in the SADRC approach. Although optimized for $1.5\:\mathrm{W}$ operation, this configuration induces persistent oscillations at all other power levels. Corresponding Allan variance curves (Fig. \ref{fig.7}\subref{fig.7b}, light dashed lines) confirm degraded mid-to-long-term stability, showing pronounced increases between $10^1\:\mathrm{s}$ and $10^3\:\mathrm{s}$ correlation times.  
	
	Conversely, the DLADRC strategy demonstrates marked improvement through its supplemental ESO. As quantified in Table \ref{table_3}, the embed of inner loop reduces 1-hour AL power instability by over 57\% compared to SADRC. Notably, the graphical results manifest superior stability performance, supported by two critical findings: (1) the uniform dark traces in Fig. \ref{fig.7}\subref{fig.7a} indicate highly consistent time-domain behavior across all power levels; and (2) the Allan variance analysis in Fig. \ref{fig.7}\subref{fig.7b} reveals remarkable stability metrics, with $<{10^{-9}} \: {\mathrm{{W}}^2}$ fluctuations observed in the short-term regime ($10^{-1}\:\mathrm{s}$--$10^0\:\mathrm{s}$) and $<{10^{-7}} \: {\mathrm{{W}}^2}$ instability maintained over longer duration ($10^1\:\mathrm{s}$--$10^3\:\mathrm{s}$). In alignment with analysis in Section \ref{sec.3a}, this robustness improvement originates from the inner-loop ESO's augmented capacity to accommodate uncertain system model's variation across diverse operating conditions.
	
	\subsubsection{Experiment E3 ---  Verification of Noise Decoupling Effect of the Cascade Outer Loop}
	\begin{figure}[!t]
		\centering
		\subfloat[(a)\hspace{3cm}(b)]
		{		
			\label{fig.8a}
			\begin{minipage}[h]{0.7\columnwidth}
				\includegraphics[width=\textwidth]{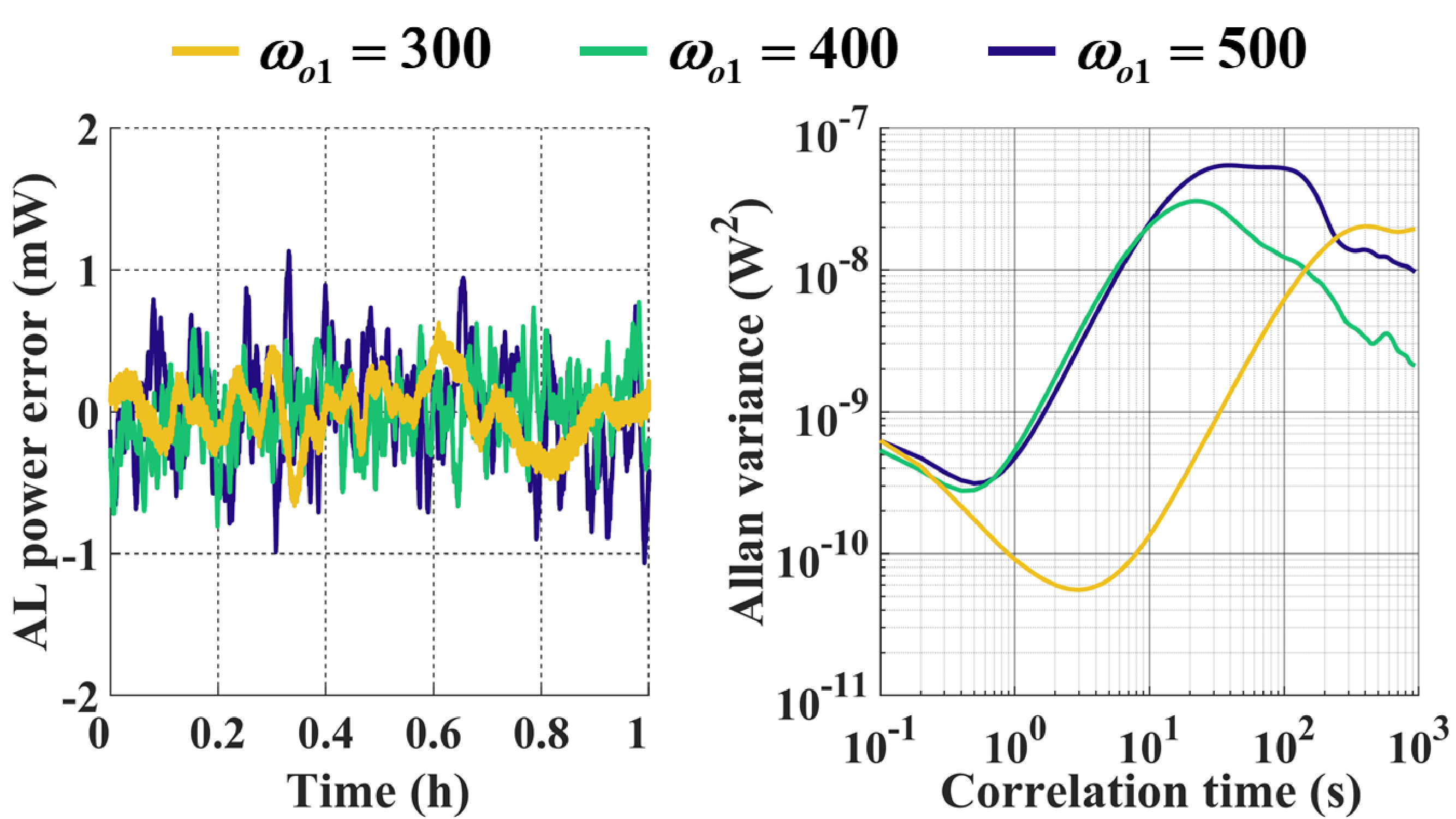}
			\end{minipage}
		}
		\subfloat[]
		{
			\label{fig.8b}
		}
		\\
		\subfloat[(c)\hspace{3cm}(d)]
		{		
			\label{fig.8c}
			\begin{minipage}[h]{0.7\columnwidth}
				\includegraphics[width=\textwidth]{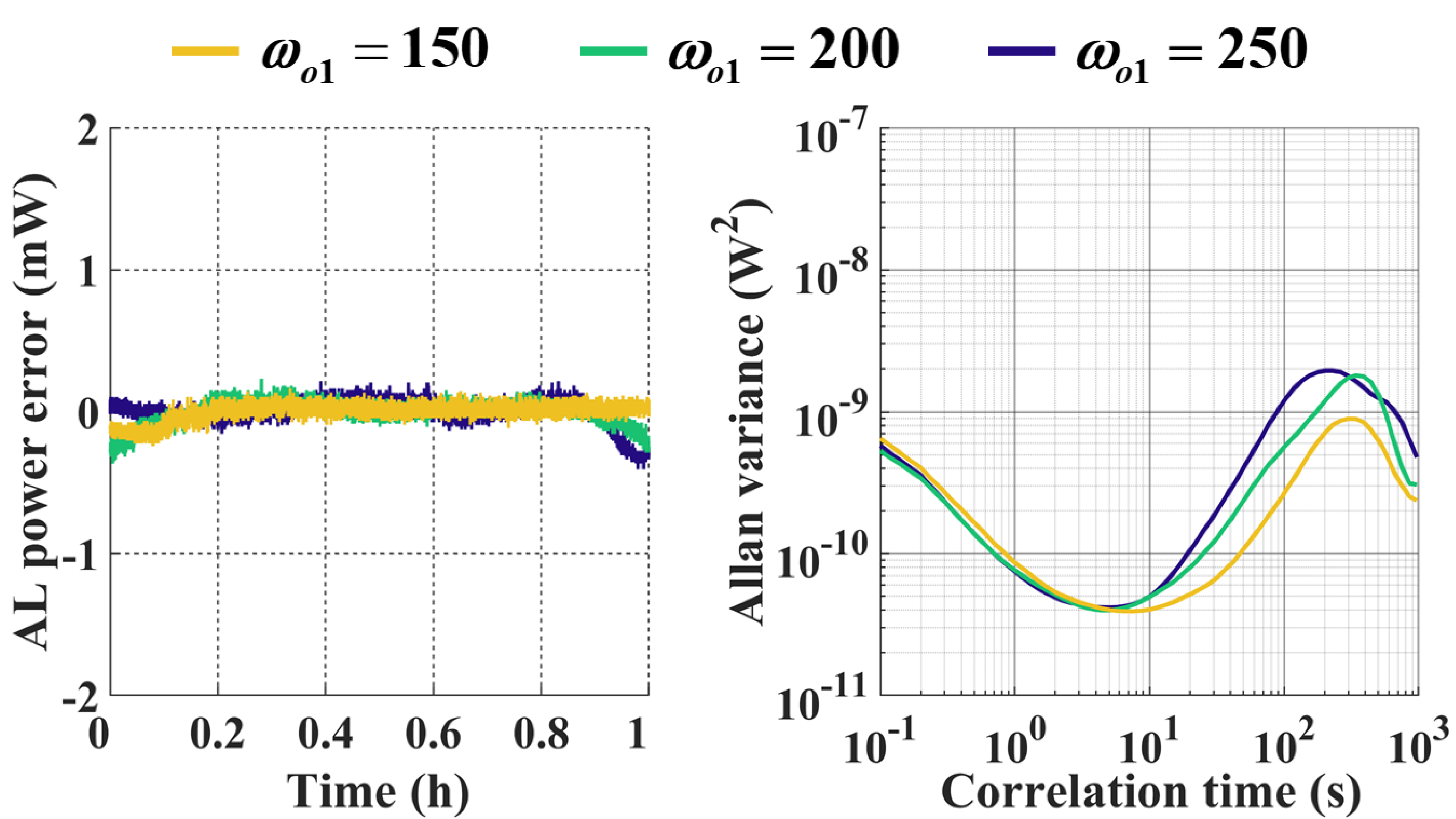}
			\end{minipage}
		}
		\subfloat[]
		{
			\label{fig.8d}
		}
		\caption{Results of E3. (a) Time domain and (b) Allan variance at $p=2$. (c) Time domain and (d) Allan variance at $p=3$.}
		\label{fig.8}
	\end{figure}
	
	Building upon the analysis of single-level outer-loop ESO ($p=1$) in E2, E3 investigates the noise decoupling performance of the cascade outer loop structure in the DLADRC strategy for cases with $p=2$ and $p=3$, under varying bandwidth configurations. To ensure consistency in observation ranges, the $p$-th level ESO is assigned a fixed bandwidth ${\omega _{op}}$. The experiment is conducted at the standard operational point of $1.5\:\mathrm{W}$.
	
	The results, presented in Fig. \ref{fig.8}, demonstrate the combined impact of cascade ESOs and bandwidth tuning on control performance. As shown by the color-matched curves in Fig. \ref{fig.8}\subref{fig.8a} and \subref{fig.8c}, increasing the cascade levels $p$ significantly improves the stability of AL power while attenuating sensor noise. Specifically, the 1-hour AL power instability is reduced to $<0.05\%$ without compromising disturbance rejection. This improvement stems from the cascade structure's ability to dilute noise content by tightening the observation range of the first-level ESO directly connected to system output.
	
	In addition to promoting stability, the cascade structure also optimizes ESO bandwidth configuration. For the $p=2$ case, Fig. \ref{fig.8}\subref{fig.8a} reveals a critical trade-off in bandwidth selection. Lower ${\omega _{o1}}$ values effectively filter high-frequency noise but risk sluggish state estimation, leading to degraded stability, as seen in the yellow curve. Conversely, higher ${\omega _{o1}}$ values enhance responsiveness but amplify noise contamination, evident in the spiked oscillations of the green and purple curves. This limitation is resolved by increasing the cascade levels to $p=3$, where the three error curves in Fig. \ref{fig.8}\subref{fig.8c} exhibit more consistent performance, indicating reduced sensitivity to ${\omega _{o1}}$ selection. The cascade structure thus relaxes bandwidth constraints while maintaining high stability.
	
	These advancements are further substantiated by Allan variance analysis. While $p=2$ struggles to balance stability across timescales through ${\omega _{o1}}$ tuning, $p=3$ maintains robust performance with negligible dependence on ${\omega _{o1}}$, as reflected in the more clustered Allan variance curves (Fig. \ref{fig.8}\subref{fig.8d} vs. Fig. \ref{fig.8}\subref{fig.8b}). Notably, long-term stability (timescales $>{10^2}\:{\mathrm{s}}$) is improved at least an order of magnitude compared to the SADRC (dashed lines in Fig. \ref{fig.7}\subref{fig.7b}), underscoring the cascade structure's superiority. In summary, the cascade outer loop delivers substantial sensor noise decoupling with bandwidth optimization, whereby resolving the inherent constraints in standard structures.
	
	Finally, Table \ref{table_4} presents a concise summary of the results from this article, by comparing the proposed DLADRC with the SADRC based on a set of criteria.
	
	\begin{table}[!t]
		\renewcommand{\arraystretch}{1.3}
		\caption{Comparison between SADRC and DLADRC}
		\centering
		\label{table_4}
		\resizebox{\columnwidth}{!}{
			\begin{tabular}{p{23mm} c c c}
				\hline\hline 
				\diagbox{Criterion}{Strategy} &
				\multicolumn{1}{c}{SADRC} &
				\multicolumn{1}{c}{DLADRC ($p=1$)} & \multicolumn{1}{c}{\pbox{20cm}{DLADRC ($p \geqslant 2$)}} \\
				\hline
				
				Design parameters  & ${\omega _{\text{c}}},\: {\omega _{o1}}$ & ${\omega _{\text{c}}},\:{\omega _{{\text{in}}}},\:{\omega _{o1}}$ & ${\omega _{\text{c}}},\: {\omega _{{\text{in}}}},\:{\omega _{o1}},\:\alpha,\:p$\\
				
				Robustness against operation variation & --- & $\nearrow$ & $\nearrow$ \\ 
				
				Noise rejection  & --- & --- & $\nearrow$ \\ 
				
				Short-term stability (Allan variance @ $10^{-1}\:\mathrm{s}$ --$10^1\:\mathrm{s}$) & --- & $\nearrow$ & $\nearrow$\\
				
				Long-term stability (Allan variance @ $10^{2}\:\mathrm{s}$ --$10^3\:\mathrm{s}$) & --- & $\nearrow$ & $\nearrow\nearrow$\\
				\hline\hline
			\end{tabular}
		}
	\end{table}
	
	\section{Conclusion}
	\label{sec.5}
	In conclusion, this study presents a novel compact AL power stabilization approach using a DLADRC strategy, while establishing a unified explicit estimation framework for both observation and control errors in the dual-loop system with experimental validation. Although requiring slightly greater parametric complexity than conventional architecture, the proposed DLADRC achieves superior stability improvements through coordination between its supplemental inner-loop ESO and cascade outer-loop ESOs. Notably, extensive experimental evaluation across the full watt-level operational range demonstrates an over $85.7\%$ reduction in 1-hour AL power instability and a tenfold decrease in Allan variance for correlation times spanning $10^1\:\mathrm{s}$--$10^3\:\mathrm{s}$. Moreover, these advancements in AL stabilization technology provide critical capabilities for large-scale OPM array optimization.
	
	
	\bibliographystyle{Bibliography/IEEEtran}
	\bibliography{Bibliography/BIB_DLADRC} 

\end{document}